\documentclass[11pt]{article}

\usepackage[sort&compress,numbers]{natbib}
\usepackage{bibentry}
\usepackage{tikz}
\usepackage{answers}
\usepackage{cellspace}
\usepackage{pdflscape}
\usepackage{amssymb}
\usepackage{adjustbox}
\usepackage{pgfplots}
\usepackage{graphicx}
\usepackage{hyperref}
\usepackage{float}
\usepackage{multicol}
\usepackage{multirow}
\usepackage{titling}
\usepackage{cancel}
\usepackage{mathrsfs}
\usepackage{geometry}
\usepackage{amsmath,amsthm,amssymb}
\usepackage[shortlabels]{enumitem}

\theoremstyle{definition}

\theoremstyle{remark}

\theoremstyle{example}

\theoremstyle{conjecture}

\newcommand{\subtitle}[1]{%
	\posttitle{%
		\par\end{center}
	\begin{center}\large#1\end{center}
	\vskip0.5em}%
}

\begin{document}
\title{A review of topological data analysis and topological deep learning in molecular sciences}
\author{JunJie Wee$^{1}$, Jian Jiang$^{1,2,\ast}$\\
\normalsize{$^{1}$Department of Mathematics, Michigan State University, East Lansing, MI 48824, USA}\\
\normalsize{$^{2}$ Research Center of Nonlinear Science, School of
Mathematics and Statistics}\\
\normalsize{Wuhan Textile University, Wuhan, 430200, P R. China}\\\\
\normalsize{$^\ast$ Address correspondences to Jian Jiang. E-mail: 
jjiang@wtu.edu.cn}
}

\date{\today}

\maketitle

\begin{abstract}

Topological Data Analysis (TDA) has emerged as a powerful framework for extracting robust, multiscale, and interpretable features from complex molecular data for artificial intelligence (AI) modeling and topological deep learning (TDL). This review provides a comprehensive overview of the development, methodologies, and applications of TDA in molecular sciences. We trace the evolution of TDA from early qualitative tools to advanced quantitative and predictive models, highlighting innovations such as persistent homology, persistent Laplacians, and topological machine learning. The paper explores TDA’s transformative impact across diverse domains, including biomolecular stability, protein–ligand interactions, drug discovery, materials science, and viral evolution. Special attention is given to recent advances in integrating TDA with machine learning and AI, enabling breakthroughs in protein engineering, solubility and toxicity prediction, and the discovery of novel materials and therapeutics. We also discuss the limitations of current TDA approaches and outline future directions, including the integration of TDA with advanced AI models and the development of new topological invariants. This review aims to serve as a foundational reference for researchers seeking to harness the power of topology in molecular science.
\end{abstract}

{\setcounter{tocdepth}{4} \tableofcontents}
\section{Introduction}

Molecular science is a broad field of scientific inquiry that focuses on studying the structure, properties, and interactions of molecules, which are the fundamental building blocks of matter, materials, chemicals, and life. It encompasses various disciplines, including chemistry, physics, biology, materials science, and nanotechnology. Molecular science seeks to understand the behavior of molecules at the atomic and molecular levels and how this behavior impacts macroscopic properties and phenomena. Overall, molecular science is a multidisciplinary field that spans multiple disciplines and explores the fundamental principles governing the behavior of molecules. It plays a crucial role in advancing scientific knowledge and technological innovation in areas such as medicine, materials science, energy, and environmental science. 

The basic method for the study of molecular science is experimentation, which has led to abundant data in the past few decades, facilitating data-driven discovery. However, data in molecular science are notoriously challenging to analyze due to their complexity, high dimensionality, multiscale,  high-order interactions, nonlinear relations, etc. For example, single-cell RNA sequence (sc-RNA seq) data are intrinsically high-dimensional, involving tens of thousands of dimensions. Human genomic data are excessively large, containing about 3 billion bases. Macromolecular structures are intricate and diverse. Molecular interactions are high-order and complex due to many-body effects and a wide range of interactions, from covalent bonds, non-covalent bonds, hydrogen bonds, van der Waals, $\pi$-$\pi$ stacking, electrostatic, to hydrophobic interactions. Conventional approaches for tackling molecular data include descriptive, inferential,  spatial, temporal, physical, Fourier, and statistical analyzes.  Many challenges for conventional molecular data analyzes can be effectively addressed by topological data analysis (TDA) \cite{carlsson2009topology}, an emerging field in data science and a new research frontier in applied sciences. 

A prominent technique in TDA is persistent homology. As a prominent technique in TDA, it combines concepts from algebraic topology and multiscale analysis to analyze data \cite{edelsbrunner2008persistent,zomorodian2004computing,pun2022persistent}. It detects complex topological invariants and patterns in data at various scales, which are not easily discernible with traditional geometric and statistical techniques. Topological invariants provide explainable representations of data \cite{xia2014persistent,xia2015fullerene,xia2023persistent} that cannot be obtained from other alternative methods. Topological invariants are often represented as persistence barcodes \cite{ghrist2008barcodes},
persistence images \cite{adams2017persistence,townsend2020representation}, persistence landscapes \cite{kovacev2016using,bubenik2017persistence}, and persistence surfaces \cite{adams2017persistence}.

However, persistent homology has several limitations, including the lack of localization, being restricted to point cloud data, and the inability to represent non-topological information. In recent years, effort has been devoted to addressing these challenges. In order to incorporate additional information in simplicial complexes, 
element-specific persistent homology \cite{cang2018integration}, multi-level persistent homology \cite{cang2018representability}, atom-specific persistent homology\cite{bramer2020atom}, electrostatic persistence \cite{cang2018representability}, and weighted persistent homology \cite{anand2020weighted,meng2020weighted,gao2021persistent} were proposed and applied in biomolecular data analysis. Additionally, persistent homology or TDA in general, inherently simplifies the data. As such, it may not be suitable for simple data where the TDA simplification may lead to the loss of essential information. Most data in molecular sciences are very complex and require a complexity reduction in their analysis. Over the years, many developments of TDA algorithms have been inspired by the needs and challenges in molecular sciences.  

Persistent cohomology was proposed to handle heterogeneous information \cite{cang2020persistent}. Wang et al. introduced persistent spectral theory \cite{wang2020persistent} to account for certain non-topological shape evolution. This spectral approach, also called persistent Laplacians \cite{memoli2022persistent, liu2023algebraic}, recovers the topological invariants of persistent homology via its harmonic spectra and offers additional non-topological information through its non-harmonic spectra. As a result, it outperforms persistent homology, as shown in a test on more than 30 datasets \cite{qiu2023persistent}.
Persistent Laplacians have been extended to many topological domains, such as cellular sheaves \cite{wei2025persistent3}, path complexes \cite{wang2023persistent}, hypergraphs \cite{liu2021hypergraph}, hyperdigraphs \cite{chen2023persistent}, and directed flag complexes \cite{zia2025persistent}. An effective software package has been developed for computing persistent topological Laplacians \cite{jones2025petls}.  
This approach was generalized to the Dirac operator in terms of quantum persistence \cite{ameneyro2024quantum}. Persistent Dirac operators have been considered for simplicial complexes \cite{ameneyro2024quantum,wee2023persistent},
path complexes \cite{suwayyid2024persistent}, and Mayer-Dirac operators \cite{suwayyid2024persistent2}. In addition to these topological spectral formulations, persistent Mayer topology was proposed to generalize the chain complex structure in algebraic topology \cite{shen2024persistent,suwayyid2024persistent2}. Moreover,  persistent interaction topology \cite{liu2023interaction, liu2025persistent} has also been proposed for point cloud data. These new formulations extend traditional algebraic topology approaches in data science. Among these new methods, persistent cohomology \cite{cang2020persistent}, persistent sheaf Laplacians \cite{wei2025persistent3}, and persistent interaction topology enable local topological analysis \cite{liu2023interaction, liu2025persistent}.  A survey of these approaches is available \cite{wei2025persistent}. 
For data on differential manifolds, the evolutionary de Rham-Hodge method \cite{chen2021evolutionary} and persistent de Rham-Hodge Laplacians \cite{su2024persistent} have been developed. These methods enable manifold topological deep learning of biomedical data \cite{liu2025manifold}. Finally, for one-dimensional curves embedded in three-dimensional space, the multiscale Gauss link integral \cite{shen2024knot}, multiscale Jones polynomials \cite{song2025multi}, and persistent Khovanov homology \cite{liu2024persistent} have been introduced. While the multiscale Gauss link integral can be easily applied, the computation of persistent Khovanov homology for knots, links, and tangles with a large number of crossings remains a challenge \cite{shen2025computing, shen2025khovanov, shen2024evolutionary}. Other methodological advances in TDA beyond persistent homology are discussed in more detail in a recent review \cite{su2025topological}.

Many data are very complex and cannot be analyzed using intuitive approaches. Therefore, TDA must be combined with machine learning (ML) to achieve its goals.  The first integration of TDA and deep neural networks, called topological deep learning (TDL), was introduced by Cang and Wei in 2017 \cite{cang2017topologynet}. TDL is an emerging paradigm in data science and a new frontier rational learning\cite{papamarkou2024position}. In the past decade, TDL has been extensively applied in molecular sciences \cite{pirashvili2018improved}. Among the most persuasive illustrations that reliably highlight the key benefits of TDL compared to traditional approaches across diverse domains are TDL's triumphs in the D3R Grand Challenges, a global annual contest focused on computer-aided drug design \cite{nguyen2019mathematical,nguyen2020mathdl}, the revelation of SARS-CoV-2 evolutionary mechanisms \cite{chen2020mutations,wang2021mechanisms}, and the precise predictions of emerging dominant SARS-CoV-2 variants BA.2 \cite{chen2022omicron} and BA.5 \cite{chen2022persistent} approximately two months ahead.  Recent advances in TDL have been reviewed \cite{papamarkou2024position}.

Molecular sciences encompass a diverse range of studies, from understanding the structure and behavior of individual molecules, investigating complex interactions within biological systems, the design of functional materials, to the discovery of effective drugs for treating diseases. The wealth of data generated in molecular sciences demands innovative tools that can effectively unravel the hidden patterns and relationships embedded in the information. 
In the past decade, TDA and TDL have demonstrated numerous successful applications across various domains in various aspects of molecular sciences. These include macromolecules, drug design and discovery, materials science, and many others. TDA and TDL have accomplished the utmost promise of uncovering hidden patterns, characterizing molecular structures,   viral evolution \cite{chen2022persistent,chen2022omicron,chen2022emerging}, and protein engineering \cite{qiu2023persistent,qiu2023artificial}. The review of the early development of TDA,  TDL, and other mathematical methods, such as differential geometry and graph theory, for biomolecular systems was given in 2020 \cite{nguyen2020review}. 

Despite of outstanding achievements in TDA applications to molecular sciences, there are obstacles in TDA methodologies that limit its further development. The current TDA-based feature engineering approaches may fall short in capturing the intricate topology inherent in molecular structures and interactions. Additionally,  as the molecular datasets continue to expand in size and complexity, the limitations of conventional analytical methods and current TDA approaches become more apparent. Moreover,  AI technologies are fast evolving. For example, there have been several groundbreaking achievements in AI and its applications to molecular science, such as  AlphaFold \cite{abramson2024accurate,wei2019protein}. 
Although there are efforts that  explore  the use of TDA in the context of AlphaFold and ChatGPT for molecular science 
 \cite{wang2023chatgpt,qiu2023persistent}, it remains an open problem as how to further position TDA along with rapidly  evolving AI technologies. These gaps and challenges   call for the integration of cutting-edge AI and new TDA methodologies that can provide a more nuanced and comprehensive understanding of molecular data. To facilitate the future development, it is imperative to conduct a timely survey  to compile the major TDA approaches with its applications in molecular sciences and to discuss remaining challenges. 

In this article, we first review the early research and development of TDA in molecular sciences that have a crucial impact on the current success of TDA and its applications. Thereafter, we highlight the most significant TDA achievements in macromolecules, drug discovery, materials science, and other fields. Lastly, we discuss an outlook for new TDA methodologies and their applications in molecular sciences. Through this review, we strive to contribute to the understanding of TDA's usages, potentials, and challenges in advancing molecular research and inspiring further exploration in this interdisciplinary field.

\section{Early research and development of TDA in molecular sciences}

\subsection{Qualitative and descriptive analysis} 
Some of the first applications of topology-related techniques to molecular sciences were qualitative and descriptive, aimed to reveal biomolecular structure and function relationship. For example, a topology-related tool, alpha complex, was applied to the anatomy of protein pockets and cavities in 1998 \cite{liang1998anatomy}. This approach has a potential for understanding protein-ligand binding. 
Another TDA tool, MAPPER, was utilized to analyzed 
the folding of a small RNA tetraloop hairpin in 2008 \cite{bowman2008structural}. The same technique was also applied to the study of biomolecular folding pathways in 2009 \cite{yao2009topological}. 

\subsection{Quantitative and predictive analysis} 

 Xia an Wei introduced TDA as a quantitative and predictive tool in 2014 \cite{xia2014persistent}. Persistent homology was, for the first time, introduced for extracting explainable molecular topological fingerprints (MTFs) in 2014 \cite{xia2014persistent}.  MTFs were employed for protein characterization, identification, and classification. This work offered both all-atom and coarse-grained representations of MTFs to shed light on the optimal cutoff distance in elastic network models for proteins and gave a quantitative modeling of protein flexibility,  predicting the optimal characteristic distance used in protein B-factor analysis. Finally, MTFs are used to characterize protein topological evolution during protein folding and quantitatively predict protein folding stability. This work reveals the topology–function relationship of proteins. 

Xia et al. introduced a persistent homology-based method to quantitatively predict fullerene stability \cite{xia2015fullerene}. They discovered that the heat of formation energy correlates with the local hexagonal cavities in small fullerenes, while the total curvature energies of fullerene isomers are linked to their sphericity, as quantified by the lengths of their long-lived Betti-2 bars. Their approach showed strong correlation coefficients between persistent homology predictions and results from quantum or curvature analysis.

One of the first TDA-based machine learning studies was due to Cang et al. in 2015 \cite{cang2015topological}. In their study, the authors investigated the potential of persistent homology as an independent tool for protein classification. They proposed a molecular topological fingerprint-based support vector machine (MTF-SVM) classifier, constructing machine learning feature vectors exclusively from protein topological fingerprints, which are topological invariants derived during the filtration process. To validate their MTF-SVM approach, they addressed four types of problems. First, they analyze protein-drug binding using the M2 channel protein of influenza A virus, achieving 96\% accuracy in distinguishing drug-bound and unbound M2 channels. Second, they assessed the classification of hemoglobin molecules in their relaxed and taut forms, obtaining approximately 80\% accuracy. Third, they performed identification of all-alpha, all-beta, and alpha-beta protein domains using 900 proteins, achieving an 85\% success rate. Finally, they applied their TDA technique to 55 protein superfamily classification tasks involving 1357 samples and 246 tasks involving 11,944 samples, attaining average accuracies of 82\% and 73\%, respectively.  
This study established computational topology as an independent and effective alternative for protein classification.

Gameiro et al.  partially elucidated the relationship between a protein's compressibility and its molecular geometric structure in 2015 \cite{gameiro2015topological}. To explore and comprehend the significant topological features of a protein, they modeled its molecule using alpha filtration, which provides multi-scale insights into the structure of its tunnels and cavities. By analyzing persistence diagrams, they derived a compressibility measure based on topological features that are indicated to be relevant by the protein's physical and chemical properties. Their primary finding demonstrates a clear linear correlation between this topological measure and the experimentally determined compressibility for most proteins where both PDB information and experimental compressibility data are available.

Other early applications of persistent homology to molecular sciences include persistent topology for cryo-EM data analysis in 2015 \cite{xia2015persistent}, and persistent homology analysis of excessive large biomolecular datasets in 2025 \cite{xia2015multiresolution}.   
Persistent homology and dynamical distances were used to analyze protein binding  in 2016 \cite{kovacev2016using}. 

\subsection{Early TDA technical developments for molecular sciences}

\paragraph{Multidimensional persistence} 
In 2015, Xia and Wei introduced  multidimensional persistence  \cite{xia2015multidimensional} and multiresolution persistent homology
\cite{xia2015multiresolution_large, xia2015multiresolution}.
They explored persistent homology for simplifying complex biomolecular data, focusing on multidimensional persistence to connect geometry and topology. They introduced pseudomultidimensional persistence derived from repeated homology filtration on high-dimensional data like molecular dynamics, and multiscale multidimensional persistence, using isotropic and anisotropic scales to form new simplicial complexes. These methods' utility, robustness, and efficiency are demonstrated in protein folding, flexibility analysis, cryoelectron microscopy denoising, and nanoparticle scale dependence. They observed topological transitions in protein folding and use Laplace–Beltrami flow to distinguish noise from molecular topological fingerprints. Multiscale persistence highlights local Betti-0 features and global Betti-1 and Betti-2 characteristics. These approaches are the early version of persistence images \cite{adams2017persistence} and are also connected to multiparameter \cite{vipond2021multiparameter,su2025topological}.

\paragraph{Object-oriented persistent homology}

Earlier applications of persistent homology before 2014 were essentially limited to qualitative data description and analysis. Additionally, persistent homology served as a passive tool for data analysis, rather than a proactive one. A new object-oriented persistent homology was proposed to fill this gap using differential geometry \cite{wang2016object}. A surface free energy functional was defined, and its minimization produces a Laplace–Beltrami operator, creating a multiscale data representation for targeted filtration. This preserved geometric features, enhancing topological persistence. A cubical complex algorithm was used to update the Laplace–Beltrami flow in the Cartesian representation. Extensive validation confirmed consistency with Euclidean distance filtration and reliability across mesh sizes. The method analyzed protein and fullerene topologies, predicting fullerene isomer stability via a model linking cavity persistence to curvature energy. Over 500 fullerene molecules verified its robustness, offering accurate predictions for ten isomer types. This work connects persistent homology with different geometry analysis and partial differential equation (PDE) modeling. 

\paragraph{Element-specific persistent homology} 

Topology offers the maximal abstraction but often loses geometric detail. Persistent homology integrates topology and multiscale analysis to capture topological invariants and  connecting complex geometry with abstract topology. However, it oversimplifies essential geometric information. A new method, element-specific persistent homology (ESPH) or multicomponent persistent homology was introduced in 2017 to preserve essential biological information during topological simplification 
\cite{cang2017analysis,cang2018integration}. 
 By partitioning the point cloud into different atom subsets, element-specific persistent homology can capture the topological features of different atom-atom interactions in proteins. For example, ESPH features generated from carbon atoms are associated with hydrophobic interactions. Similarly, interactions between nitrogen and oxygen atoms correlate to hydrophilic interactions and/or hydrogen bonds. By combining ESPH with machine learning, a robust framework for macromolecular analysis was developed to capture physical, chemical, and biological information and interactions in macromolecular systems. For the prediction of protein stability changes upon mutation, this topological approach outperformed the state-of-the-art competing methods \cite{cang2017analysis}. Further testing on two large showed this ESPH-based machine-learning approach surpasses existing methods in protein-ligand binding affinity predictions \cite{cang2018integration}. ESPH uncovered protein-ligand binding mechanisms unattainable by conventional techniques, revealing hydrophobic interactions extending 40\AA~ from the binding site, significantly impacting drug and protein design. The essential idea of ESPH has served as a cornerstone for the TDA of complex data in the past decade.

\paragraph{Topological deep learning}

Topological deep learning (TDL), first introduced by Cang and Wei in 2017 \cite{cang2017topologynet}, is a fast-growing field that leverages topological features to enhance the understanding and development of deep learning models. Due to its explainability, TDL represents a new frontier in relational learning \cite{papamarkou2024position}. By integrating topological concepts, TDL complements graph representation learning and geometric deep learning, making it a promising approach for various machine learning applications. The original development of TDL, also called topologyNet, was motivated by the need to predict large datasets in protein-ligand binding affinities and protein stability changes upon mutation. This work also utilized element-specific persistent homology in feature generation. The element-specific method characterizes different atom-atom interactions instead of directly applying persistent homology on the entire point cloud of a protein-ligand complex. The vectorization of these element-specific features was used in various deep neural networks, including  convolutional neural networks (CNNs) and multi-task neural networks (MTNN) for biomolecular property predictions. Due to the interpretablility of persistent homology. These methods were some of the first interpretable neural networks (INN).  The results of this approach were the state-of-the-art in the field. TDL is a new paradigm in data science and machine learning \cite{zia2024topological}.  

\paragraph{Multi-level persistent homology and electrostatic persistence}

Cang et al. proposed various methods to enhance persistent homology for representing biomolecular data in machine learning and deep learning applications \cite{cang2018representability}. One challenge is that biomolecules and small molecules exhibit multiscale interactions, including electronic, atomic, functional group, residue, domain, intramolecular, and intermolecular scales. These interactions determine critical physical, chemical, and biological properties. A simplistic filtration approach cannot adequately capture these interactions. To address this, Cang et al. introduced multilevel persistent homology to extract interactions at appropriate spatial scales. Additionally, the authors developed electrostatic persistence to represent, characterize, and describe small molecules and biomolecular complexes. Molecular electrostatics involves the electrostatic forces that govern the structure, function, dynamics, and interactions of molecules such as ligands, drugs, proteins, and DNA, influenced by charged molecules, amino acids, and the surrounding ionic environment. Electrostatics plays a crucial role in molecular science. Electrostatic persistence captures molecular electrostatics through charge embedding schemes, both utilizing atomic partial charges computed via physical models. Consequently, electrostatic persistence enables the development of physics-informed neural networks (PINNs).

\section{Macromolecules and interactions}

\subsection{Biomolecular stability and solubility}
Biomolecular stability refers to the ability for a biomolecule, a protein or DNA/RNA to preserve its configuration despite alterations in environmental conditions like temperature, pH, and the composition of the solvent. 
Solubility indicates the capacity of a biomolecule to remain dissolved in solution under physiological or formulation conditions, critically influencing its stability and bioavailability. Particularly, biomolecular mutations can significantly influence the stability and solubility of a protein or DNA/RNA, which results in altered biomolecular functions and leads to various diseases. Over the past decades, TDA has made significant contributions to the understanding and prediction of biomolecular stability and solubility, providing a unique perspective on the complex relationships within biomolecular structures \cite{xia2014persistent}. By representing biomolecular structures as simplicial complexes, where amino acids are nodes, and edges or higher-dimensional simplices represent interactions, such as hydrogen bonds or van der Waals forces, TDA captures the topological features of the biomolecule, offering a more holistic view beyond traditional methods that focus on geometric or energetic considerations. Various early works of TDA have introduced persistent homology for mathematical modeling of biomolecular structures, providing the necessary application of TDA into biomolecular function and dynamics. 

In the initial applications of TDA for molecular sciences, persistent homology was utilized to study the stability of fundamental biomolecular structures, including fullerene and carboranes and carbon isomers \cite{chen2020persistent,xia2015fullerene}. The goal was to gain a deeper understanding of how topological fingerprints are being influenced by the biomolecular structural changes. For instance, persistent homology has been analyzed on large biomolecular data \cite{xia2015multiresolution_large,xia2015multiresolution,xia2015multidimensional} and efficiently resolved ill-posed inverse problems in cryo-EM structure determination \cite{xia2015persistent}. Further, persistent homology could provide the topological and geometric information to calculate a topological compressibility for proteins which is well related to the structure stability of proteins \cite{gameiro2015topological}. Persistent homology contributed greatly in protein folding analysis by extracting topological features through protein unfolding process, allowing researchers to construct models to understand topology-function relationship between topological features and protein stability \cite{verovvsek2016extended,harvey2014collaborative,mirebrahimi2019persistent}. Additionally, 
gradient boosting trees using integrated persistent Laplacian and pre-trained transformer-based features were developed to predict mutation-induced protein solubility changes \cite{wee2024integration}. These contributions solidified TDA as a robust and powerful tool in structure analysis of biomolecular data.

\subsection{Biomolecular flexibility}
The protein B-factor, also known as the temperature factor or Debye–Waller factor, quantifies the atomic displacement or structural flexibility within a crystal structure, providing insights into local protein dynamics and conformational stability. Accurate prediction of protein B-factors is an important and meaningful metric in understanding the protein structure, flexibility, and function. The TDA-based approaches offer a robust framework for predicting protein B-factors by capturing intrinsic structural features and multi-scale geometric and topological patterns, enabling noise-resistant characterization of flexibility and dynamic behavior beyond conventional coordinate-based methods.

Mathematically, we need localized  models for B-factor predictions. However, persistent homology is global because its topological invariants are typically for the whole dataset.  Recently, new approaches such as weighted persistent homology \cite{pun2020weighted}, atom-specific persistent homology \cite{bramer2020atom}, and evolutionary homology \cite{cang2020evolutionary} for protein flexibility analysis.  
In 2025, Hayes et al. proposed the persistent sheaf Laplacian (PSL), a new effective tool in TDA, to model and analyze protein flexibility \cite{hayes2025persistent}. By representing the local topology and geometry of protein atoms through the multiscale harmonic and nonharmonic spectra of PSLs, the proposed model effectively captures protein flexibility and provides accurate, robust predictions of protein B-factors.

\subsection{Biomolecular interactions}
Protein-protein interactions (PPIs) are fundamental to cellular signaling, structural organization, and complex formation. TDA made significant contributions on PPIs studies by revealing hidden network structures, identifying functional protein modules, and capturing higher-order connectivity patterns compared to traditional graph-based methods.

In the past few years, researchers have developed TDA-based AI models for predicting the changes of PPI binding affinities induced by mutations. One of the first TDA-based AI model developed for predicting the changes of PPI binding affinities induced by mutations is TopNetTree\cite{wang2020topology}, which integrates persistent homology-based features with a new deep learning algorithm called NetTree that takes advantage of convolutional neural networks and gradient-boosting trees. Thereafter, TDA-based AI models that use persistent Hom complex and persistent Tor-algebra were developed and similarly outperformed existing state-of-the-art models in predicting the changes of PPI binding affinities induced by mutations \cite{liu2022hom,liu2022persistent}. Persistent homology and molecular dynamics were also employed to investigate amino acid mutation-induced structural changes for protein-protein interactions \cite{koseki2023topological}. Additionally, Wee and Xia proposed persistent spectral (PerSpect) based PPI representation and featurization and PerSpect based ensemble learning models for PPI binding affinity prediction \cite{wee2022persistent}.

Unlike the above problems that involve mutation, the direct prediction of protein-protein binding affinity is also important\cite{shen2023svsbi}. Xu et al. recently proposed the Persistent Laplacian Decision Tree (PLD-Tree), a new approach for the direct prediction of protein–protein binding affinities\cite{xu2024pld}. This method targets protein chains at binding interfaces, utilizing persistent Laplacian to capture topological invariants that reflect key inter-protein interactions. These topological features are enriched by integrating sequence-based data from evolutionary scale modeling (ESM) derived from a large language model.

Peptide–protein interactions are crucial to biological processes such as signal transduction, enzymatic regulation, and immune responses, underscoring the importance of precise structural modeling for therapeutic innovation. Dai et al. introduced TopoDockQ, a TDL approach that accurately predicts DockQ scores (p-DockQ) to enhance model selection. Compared with AlphaFold2’s built-in confidence score, TopoDockQ reduces false positives by at least 42\% and increases precision by 6.7\% across diverse benchmarks \cite{dai2025topological}. This advancement expands the possibilities for peptide engineering, particularly in developing therapeutics with customized biochemical properties, enabling more accurate and flexible peptide design.

\subsection{Viral evolution and analysis }
Viral evolution involves genetic mutation, selection, and adaptation that drive viral diversity and pathogenicity. The Severe Acute Respiratory Syndrome Coronavirus 2 (SARS-CoV-2) caused a global COVID-19 pandemic since late 2019, which evolved into numerous variants that have led to several waves of COVID-19 infections. SARS-CoV-2 uses mutations to improve its evolutionary adaptability. TDA has advanced the study of viral evolution by uncovering hidden evolutionary structures, identifying recombination events, and capturing high-dimensional genomic patterns beyond traditional phylogenetic methods. Additionally, TDA based methods and AI could offer an accurate and efficient alternative to the experimental determination of viral infectivity, vaccines and antibody discovery \cite{chen2022mathematical}, and the binding free energy change after mutation \cite{chen2020mutations,liu2025machine}.

For instance, TDA can be applied to model the interactions between COVID-19 spike protein in the receptor-binding domain (RBD) and the host’s angiotensin-converting enzyme 2 (ACE2) \cite{wei2022topological}. Topological AI models in protein-ligand interactions has also been extended to study drug target interactions with SARS-CoV-2 main protease and drug resistance with Pfizer’s drug PAXLOVID \cite{nguyen2020unveiling,chen2024drug}. A review about SARS-CoV-2 modeling, simulations, and predictions focused on its molecular-level methodologies from the aspects of biophysics, mathematical approaches, and ML, including deep learning, bioinformatics, and cheminformatics in the applications of SARS-CoV-2 is detailed in \cite{gao2022methodology}. Wee and Wei proposed an AlphaFold3-assisted multi-task topological Laplacian model to enhance the prediction of deep mutational scanning and binding free energy changes upon virus mutations \cite{wee2025rapid}.

Topological AI models such as TopNetmAb were carried over to predict the binding free energy changes for SARS-CoV-2 RBD-ACE2 and RBD-antibody complexes due to RBD mutations \cite{chen2021prediction,chen2021revealing, bi2023multiscale}. TopNetmAb is also used to analyze how the RBD mutations on the Omicron variant affects the viral infectivity and efficacy of existing vaccines and antibody drugs \cite{chen2022omicron}.

Other TDA-based AI model, TopNetTree was developed in the study of viral mutation \cite{chen2021sars}. Apart from the 8338 PPI entries which TopNetTree is trained on, more training data related to SARS-CoV-2 was incorporated in order to improve the overall model performances. TopNetTree was also used to analyze the SARS-CoV-2 mutations in the United States, which revealed two of four SARS-CoV-2 substrains in the United States become potentially more infectious\cite{wang2021analysis}. TopNetTree and TopNetmAb are applied in antibody studies, which allowed researchers to design and analyze COVID-19 antibody candidates by integrating TDA and AI \cite{chen2021review}.

Extending from persistent homology, a new model TopLapNetGBT which integrates persistent Laplacian with deep learning, improves the performance for predicting mutation-induced PPI binding free energy changes \cite{chen2022persistent}. More importantly, this model successfully predicted Omicron BA.4 and BA.5 as dominant variants before WHO officially announced in June 2022. Similarly, TopLapNet was applied to determine emerging dominant SARS-CoV-2 variants in 2022 \cite{chen2022emerging}. Additionally, Chen et al. proposed a topological deep learning (TDL) paradigm to facilitate in silico deep mutational scanning \cite{chen2023topological}. Persistent topological Laplacians were proposed to study the protein structures of the SARS-CoV-2 spike receptor binding domain \cite{wei2023persistent2}.

\subsection{Protein engineering}

Protein engineering designs and optimizes proteins for desired functions by modifying sequence, structure, or dynamics. The use of accumulated protein databases and ML models, especially those utilizing natural language processing (NLP), have significantly accelerated the pace of protein engineering in recent years. Advancements in TDA and AI-driven protein structure prediction tools like AlphaFold2 have enabled the development of more potent strategies for protein engineering that are guided by ML and based on structure. For instance, Xia et  al. integrated TDA with graph neural networks and AlphaFold3 for protein complex structure interface quality assessment \cite{han2025topoqa}.

Protein engineering traditionally relies on prior structural and functional knowledge, yet the vast mutational space ($20^N$) limits exhaustive exploration. While directed evolution offers breakthroughs, experimental scanning remains constrained. TDA overcomes these challenges by capturing global and local structural features in a model-free manner, enabling efficient navigation of mutational landscapes. By integrating TDA with protein engineering, researchers can identify functionally relevant variants more systematically, thus enhancing predictive power and accelerating the discovery of proteins with optimized properties.

In recent times, topological AI models have emerged as new approaches to directed evolution and protein engineering. 
Persistent Laplacian-based structural features and two auxiliary sequence embeddings are integrated with AI to capture mutation-induced topological invariant, shape evolution and sequence disparity in the protein fitness landscape \cite{qiu2023persistent}. Here, both structure and sequence-based embeddings are utilized in this ML-based protein engineering approach. It is impeccable to see that deeper ML models and emerging large-scale deep mutational scanning databases will continue to further enhance the model performance in protein engineering. A review of TDA applications in protein engineering was given by Qiu and Wei \cite{qiu2023artificial}.

\subsection{Molecular dynamics simulation}

Molecular dynamics (MD) simulations provide atomistic insights into biomolecular motions and interactions over time. Traditional MD methods often struggle with limited sampling efficiency and difficulty capturing rare events on biologically relevant timescales. TDA enhances MD by capturing global geometric and topological features, enabling robust detection of conformational states, transition pathways, and hidden collective dynamics beyond traditional coordinate-based metrics.

The first application of persistent homology to protein folding was due to Xia and Wei in 2014 \cite{xia2014persistent}. These authors studied protein folding and unfolding using molecular topological fingerprints (MTFs) to characterize protein topological evolution during protein folding and quantitatively predict the protein folding stability. They found an excellent consistence between MTF prediction and molecular dynamics simulation. 

Ichinomiya et al. used persistent homology to analyze protein folding 
\cite{ichinomiya2020protein,ichinomiya2022topological}. Recently,  Arango et  al. demonstrated that a persistent homology-based TDA approach, integrated with deep learning, outperformed traditional order parameters by capturing both local and global lipid structural features to robustly predict membrane organization across temperatures \cite{arango2023topological}. In 2024, they introduced a persistent homology–based topological learning framework that, combined with attention networks, enabled multiscale prediction of lipid effective temperatures by capturing both local and global structural features beyond traditional metrics \cite{arango2024topological}.

\section{Drug discovery}

\subsection{Drug target identification}
 
Drug target identification is fundamental in drug discovery as it defines the molecular basis of disease intervention, enabling the rational design of therapeutics with higher specificity and efficacy. For instance, Tola et al. identified potential inhibitor compounds for methylcitrate dehydratase by utilizing multiparameter persistence based on persistent homology \cite{tola2024identification}.
One limitation in traditional persistent homology used in target identification is that it treats all atoms indiscriminately. Hence, persistent path topology (PPT) was proposed to characterize molecules by incorporating its embedding with element types into its topological analysis \cite{chen2023path}. Furthermore, angle-based filtration PPT is also proposed to complement the existing distance-based filtration PPT. The benefit of PPT allows researchers to handle molecular structures without the need for techniques like element specific \cite{cang2017analysis} or persistent cohomology \cite{cang2020persistent}. With this benefit, researchers introduced topological perturbation analysis (TPA) as the inaugural technique of PPT for the analysis of biological networks \cite{chen2023path}. 

TPA has the unique capability to identify crucial nodes within intricate networks and holds immense potential for applications in various biological networks such as gene regulatory networks \cite{du2024multiscale}, protein-protein interaction networks, signaling networks, metabolic networks, neuronal networks, DNA-DNA-chromatin networks, and transcriptomic networks. This work introduced a fresh perspective to drug target discovery. PPT heralds a new era for future advancements in biological networks, showing promise for drug target identification, discovery of gene motifs, directed evolution, protein engineering, and omics in general.

\subsection{Virtual screening }

Virtual screening is a computational strategy in drug discovery that systematically evaluates large chemical libraries to identify potential bioactive compounds against specific targets. Traditional approaches often rely on molecular docking or similarity-based methods, which may overlook complex structural or functional relationships. By integrating TDA, virtual screening can capture hidden geometric and topological features of molecular space, enabling more robust discrimination between active and inactive compounds. This topological perspective enhances predictive power, improves generalization across diverse chemical scaffolds, and facilitates the identification of novel drug candidates beyond conventional chemical similarity boundaries. Additionally, virtual screening through AI learning models significantly reduces the time and cost as compared to traditional drug discovery methods \cite{keller2018phos}.

Due to the success of TDA-based deep learning models in the D3R Grand Challenges \cite{nguyen2020mathdl, nguyen2019mathematical}, researchers successfully applied TDA to further revolutionize virtual screening. Several innovative TDA-based approaches have been proposed to address challenges in identifying novel therapeutics. For example, topological Laplacian was integrated with AI learning models in virtual screening of DrugBank database for hERG blockers \cite{feng2023virtual}. Zhu et al. built a topology-inferred drug addiction learning model for virtual screening of drug addiction data by integrating multiscale topological Laplaians, deep bidirectional transformer, and ensemble-assisted neural networks \cite{zhu2023tidal}. Additionally, Keller et al. developed a persistent homology-based virtual screening tool for ligand screening \cite{keller2018phos}.

\subsection{Protein-ligand binding prediction}
The success of TDA has greatly contributed to the field of protein-ligand interactions \cite{cang2017topological,liu2023persistent,liu2021persistent}. Protein-ligand interactions is significant key area to tackle in order to advance drug design and discovery development. One key problem lies in the molecular representation and featurisation of protein-ligand interactions, which is important to build successful advanced mathematical AI models \cite{jiang2022molecular,townsend2020representation}. Extracting the important and essential topological and geometric information in molecular structures generates concise and rich topological descriptors in TDA-based AI models. TDA-based AI models have leveraged from algebraic topology, differential geometry, and algebraic graph theory to construct effective representation of biomolecular systems \cite{cang2018representability,nguyen2020review, arsuaga2012topological}. When integrated with ML or deep learning algorithms, the TDA approach applied in these advanced mathematical models have contributed to tremendous success in protein-ligand binding affinity prediction \cite{liu2022dowker,long2023predicting}. Over the years, several TDA-based deep learning models have outperformed numerous existing state-of-the-art traditional molecular descriptor ML models. \cite{nguyen2019agl,meng2021persistent,liu2021neighborhood}.

In 2015, TDA was first integrated with ML algorithms for protein classification \cite{cang2015topological}. One of the first topological deep learning model developed was in \cite{cang2017topologynet}, where TopologyNet was used to predict protein-ligand binding affinities, mutation induced globular protein folding free energy changes, and mutation induced membrane protein folding free energy changes.  

Recently, a series of tools based on TDA were developed for binding affinities prediction. For example, Feng et al. introduced commutative algebra ML for the affinity predictions of protein-ligand binding and metalloprotein-ligand binding \cite{feng2025caml}.  Zia et al. presented the persistent directed flag Laplacian (PDFL), which incorporates directed flag complexes to account for edges with directionality originated from polarization, gene regulation, heterogeneous interactions, etc. They found the proposed PDFL model outperformed competitors in protein-ligand binding affinity predictions \cite{zia2025persistent}. Feng et al. developed persistent Mayer homology (PMH) theory based on the standard homology theory, which was validated protein-ligand datasets, including PDBbind-v2007, PDBbind-v2013, and PDBbind-v2016 \cite{feng2025mayer}. Additionally, a series of advanced mathematical methods have been proposed and used on binding affinity data prediction, including persistent Hodge Laplacian learning algorithm \cite{su2024persistent}, join persistent homology \cite{wang2025join}, a novel topological ML model (TopoML) \cite{liu2025topological}, persistent Laplacian decision tree \cite{xu2024pld}, knot data analysis \cite{shen2024knot}, and multitask-topological Laplacian \cite{wee2024evaluation} etc.

The noteworthy success of TDA-based deep learning models in the D3R Grand Challenges underscores the effectiveness of this approach in real-world drug design scenarios. These models leverage TDA's capacity to extract and quantify persistent topological features, allowing for accurate predictions of molecular interactions. The ability of TDA-based AI models to outperform other methods in such competitive settings attests to the practical impact and potential transformative role of TDA in drug discovery. 

The emerging advancements in AI has also further accelerated TDA's contributions in protein-ligand interaction. From natural language processing to transformer architectures and foundational models like ChatGPT, these emerging models after pre-training on large-scale and huge amounts of databases with unlabeled data, serve as powerful solutions to boost TDA-based AI models further in protein-ligand interaction. Recently, topological transformers have demonstrated its capabilities by converting 3D protein-ligand complexes into topological sequences, thereby facilitating the application of sophisticated large language models for the analysis of protein-ligand interactions \cite{chen2024topo,chen2024drug}. In particular, the transformer has also demonstrated exceptional efficacy in predicting binding affinity tasks across a range of benchmarks, including those specifically related to SARS-CoV-2 \cite{chen2024drug}. It evaluated the influence of virus mutations on the effectiveness of drugs, providing vital understanding into potential drug resistance. Additionally, Chen et al. presented a topological transformer integrating persistent topological hyperdigraph Laplacian and transformer models for
protein-ligand interaction predictions \cite{chen2024multiscale}.

\subsection{Drug repurposing}
Another important area in drug discovery is topological AI-based drug repurposing, where topological AI models were developed to identify existing drugs repurposed for drug addiction treatment \cite{zhu2023tidal}. This can significantly reduce the drug development process, as existing drugs have already undergone extensive safety testing. In particular, antibiotic drug discovery has also garnered huge interest in drug discovery communities as topological AI models can also be developed to identify existing drugs for antibiotic resistance \cite{tarin2023computer}. This study applied a topological structure–activity data analysis model to repurpose FDA-approved drugs as candidate antimicrobials against Escherichia coli. By comparing topological signatures of drug–target interactions and E. coli proteins, the approach recovered known antibiotics and nominated diverse drug classes (e.g., antitumor agents, antihistamines, hypoglycemics) as potential antimicrobials. Cross-species topological similarities implied broader-spectrum activity and novel targets. This work demonstrated TDA-driven virtual screening as a rapid, mathematically grounded strategy for drug repurposing to address antimicrobial resistance.

Furthermore, in 2024, Du et al. proposed a TDA-enhanced method, persistent spectral theory for topological differentiation of PPI networks from differentially expressed gene data and identified three pivotal molecular targets for antiaddiction drug repurposing from DrugBank \cite{du2024multiscale}. Using topological Laplacians and algebraic-graph embeddings, Feng et al. applied TDA-integrated ML to screen DrugBank for hERG cardiotoxicity, identifying 227 potential blockers out of the 8641 DurgBank compounds and demonstrating TDA’s utility for safer drug repurposing \cite{feng2023virtual}.

\subsection{Solubility prediction}
Solubility prediction plays an important role in drug discovery by enabling early assessment of drug candidates’ developability, guiding lead optimization, and reducing late-stage attrition due to poor pharmacokinetic properties. The prediction accuracy depends crucially on molecular descriptors which are typically derived from a theoretical understanding of the chemistry and physics of small molecules. Integrating TDA-based method with ML algorithm has contributed to good prediction of aqueuous solubility with comprehensive molecular representation. 

Recently, Wu et al. proposed element specific persistent homology combined with multi-task deep neural networks for simultaneous predictions of partition coefficient and aqueous solubility, which provided multiscale and multicomponent topological invariants to describe the molecular properties and achieved some of most accurate predictions of aqueous solubility \cite{wu2018topp}. Additionally, Ehiro generated molecular descriptors from Morgan fingerprint using persistent homology and improved the prediction accuracy on solvation free energy and water solubility datasets \cite{ehiro2024descriptor}. Dong et al. proposed a modality-adaptive method based on an improved multi-objective optimization algorithm for molecular property prediction including water solubility, which integrated TDA, the teacher learning mechanism and graph centrality measures \cite{dong2025exploring}.

\subsection{Toxicity prediction}
Toxicity prediction on drug candidates is one of key procedure of drug discovery. TDA acts as a transformative tool in small molecular data challenges of toxicity datasets\cite{dou2023machine}. TDA combined with deep learning approaches can offer valuable insights into toxicity prediction. The ability of TDA to discern complex and subtle patterns in molecular data enhances the accuracy of toxicity predictions, providing a holistic view of the underlying mechanisms. This innovative approach in toxicity prediction not only aids in the early identification of potentially harmful compounds but also contributes to the development of safer and more effective drugs through a deeper comprehension of molecular interactions and their implications for toxicity. 
 
In order to enhance the prediction of small quantitative toxicity datasets, multitask deep learning models were integrated with TDA to learn multiple toxicity tasks simultaneously and exploit commonalities as well as differences across different tasks \cite{wu2018quantitative}. 
The integration of algebraic graph representations and bidirectional transformer-based embeddings with a variety of ML algorithms, including decision trees, multitask learning, and deep neural networks have produced great performance across eight molecular datasets, involving quantitative toxicity datasets\cite{chen2021algebraic}. Recently, Rong et al. proposed a topological fusion network leveraging TDA to capture multi-scale topological features and their method outperformed the state-of-the-art method by 2.4\% on ClinTox dataset for classification task \cite{rong2025topological}. Other TDA-based methods have been developed for toxicity predictions, including nanoparticle toxicology assessments \cite{offroy2025toxicity}.

\section{Materials science}
\subsection{Crystalline materials}

Progress in the field of materials science is often gradual and demanding, posing a significant challenge in keeping up with the growing need for material characterization. When dealing with intricate datasets in crystalline materials, a crucial challenge arises in determining methods to create low-dimensional representations for input crystal structures while consisting of valuable chemical insights.  One of the early applications involves the use of persistent homology to analyze defects in crystal structures \cite{grottel2011topological,stellhorn2020structure, hong2019medium}.  Persistent homology was also applied to lithium cluster structure prediction \cite{chen2020topology} and two-dimensional network analysis \cite{morley2021persistent}.

In recent years, high-throughput computational methods such as density functional theory (DFT) are employed to predict the properties of both experimental and hypothetical inorganic compounds. Although experimental and computer simulation methods have contributed to a vast amount of high-quality open databases, such methods still lack efficiency and remain expensive for heavier elements, strongly correlated electrons, and large molecules. Furthermore, DFT is not well-suited for large and diverse material datasets. As such, TDA-based descriptors have been proposed to serve as powerful tools for high-throughput screening of materials with specific topological features. These descriptors enable the identification of materials with desired properties, guiding researchers in the design and discovery of novel crystalline materials. For instance, atom-specific persistent homology (ASPH)-based ML models were developed and proved to achieve highly accurate predictions of DFT-calculated formation energy \cite{jiang2021topological}. 

Comprising metal ions or clusters linked to organic ligands, metal–organic frameworks (MOFs) represent a distinct category of crystalline materials that self-assemble into porous structures with exceptional tunability. In 2021, persistent homology  was applied to the embeddings of MOFs
\cite{krishnapriyan2021machine}.  It automatically encapsulates geometric and chemical information directly from the material system. 
In 2025, Chen et al. introduced a TDA-based category-specific topological learning for robust material property prediction \cite{chen2025category}. In 2023, Shekhar and Chowdhury introduced a feature-based representation of materials using tools from TDA for prediction of hydrogen storage in MOFs \cite{shekhar2024topological}. Additionally, Yang et al. developed a topological descriptor based on TDA, which was combined with the extreme gradient boosting algorithm to predict the adsorption performance of MOFs \cite{yang2024topological}. 

\subsection{Solar energy materials}

Recently, materials such as organic-inorganic halide perovskites (OIHPs) have gained significant attention in the field of solar cells due to their remarkable cost-effective photovoltaic capabilities, placing them in competition with widely used silicon solar cells. Although OIHP solar cells have remarkable performance, many high-efficiency OIHP solar cells contain lead and face challenges due to its poor stability under ambient environment. Consequently, this sparks an urgent interest to discover new perovskite variants with enhanced properties to overcome these limitations. 

In 2022, TDA-based approaches such as persistent homology and persistent Ricci curvature were integrated with ML to better understand and predict the formation energy and bandgap values of OHIP materials \cite{anand2022topological}.  Formation energy and bandgap are important material properties for understanding the conductive and insulating behavior of OIHP materials, crucial for optimizing their performance in solar cells. Persistent homology and persistent Ricci curvature-based descriptors also demonstrated effective classification over different OIHP crystal configurations and different atom types. Additionally, persistent homology was  applied to classify high-entropy alloy datasets containing body-centered cubic (BCC) and face-centered cubic (FCC) crystal structures \cite{spannaus2021materials}.

\subsection{Material property prediction}
Understanding the structure-function relationships in materials is the key point of material discovery. Compared with traditional experimental approaches, TDA-based methods presented their advantages in computational efficiency, uncovering structure–property correlations and providing physical insights into material behavior \cite{zheng2025active}. For instance, lithium superionic conductors (LSICs) are vital for next-gen solid-state batteries but hard to discover due to vast chemical space, limited data and complex structure-function links. In 2025, Chen et al. presented a TDA-based  multiscale topological learning framework integrating algebraic topology and unsupervised learning, extracting features and using screening metrics. They found 14 new LSICs, accelerating LSIC identification and aiding material discovery \cite{chen2025superionic}. 

Enhancing the accuracy of energy predictions in multi-lithium-atom systems is essential for optimizing atom properties and benefiting material design. Chen et al. explored the application of persistent topological Laplacian, a TDA-based method to effectively capture the intrinsic properties of many-body interactions, accelerating the discovery of new materials and enhancing the efficiency of material development \cite{chen2025enhancing}. They also introduced a TDA-based framework for extracting structural features of materials and achieved up to 55\% reduction in prediction error for defect-sensitive properties \cite{wang2025structural}. 

Simulating how molecular building blocks self-assemble into functional complexes constitutes a key research focus in materials science. Spirandelli et al. presented a long-range topological potential, measured through weighted total persistence, which they integrated with the morphometric approach to solvation-free energy \cite{spirandelli2025topological}.  Additionally, persistent multi-cover method was introduced for polymer property prediction integrating with gradient boosting tree algorithm \cite{zhang2024multi}.

\subsection{Other materials}
Persistent homology has also been applied to many other materials studies,   such as polymeric material \cite{higashi2025feature}, nano structures\cite{xia2015fullerene}, nanoporous materials \cite{krishnapriyan2020topological,fang2025leveraging}, and  soft matter \cite{landuzzi2020persistence,shimizu2021higher,membrillo2022tracking}. Persistent homology was applied to give a better understanding of the thermal stability \cite{christensen_medium-range_2023}, diffraction patterns \cite{onodera2019understanding}, hierarchical structure \cite{hiraoka2016hierarchical,minamitani2023persistent}, medium-range order of amorphous materials \cite{sorensen2020revealing, nakamura2015persistent}. For example, it was used to extract topological features from the microphase-separated structures of polymeric materials, which were successfully utilized to predict proton conductivity of polymeric materials \cite{higashi2025feature}. 

Persistent homology was combined with molecular dynamics to analyze the changes in topological information of liquid structures in \cite{sasaki2018liquid}. Additionally, persistent homology-based ML models were also developed for lithium cluster structure prediction in \cite{chen2020topology}. Software tools like HomCloud have been developed specifically for researchers who are interested in applying persistent homology and inverse analysis in the field of materials research \cite{obayashi2022persistent}.

Recently, the theory of path homology has been employed to understand the role of mirror-symmetric sublattices that obstruct the creation of periodic unit cells in amorphous materials \cite{chen2023path}. A concise review which encapsulates the developments of TDA-based ML models for material sciences is detailed in \cite{zheng2023application}. 

Additionally, representations of energy landscapes by sublevelset persistent homology were presented with n-alkanes as an example
\cite{mirth2021representations}. The shape of data and molecular representation in chemistry were discussed in relation to persistent homology analysis. \cite{townsend2020representation,bilsky2025understanding}.

\section{Other applications}

\subsection{Topological sequence analysis}
Sequence data, encompassing DNA, RNA, and protein sequences, possesses complex, multi-scale structures that present considerable challenges to conventional analysis methods that rely on alignment or purely statistical representations. While TDA has achieved increasing success in capturing the global features of complex data \cite{arsuaga2012topological,chan2013topology}, its application to sequence data is still relatively underexplored.

Hozumi and Wei introduced a k-mer topology approach for topological sequence analysis (TSA) of genomes
\cite{ hozumi2024revealing}. This method utilizes persistent homology and/or persistent Laplacian to characterize genome sequences for variant detection, species classification, and phylogenic analysis. It outperforms state-of-the-art methods. Recently, Liu et al. proposed two approaches to generalize the TDA of genomes. One approach is category theory-based TSA, which treats a sequence as a resolution category, capturing its hierarchical structure through a categorical construction \cite{liu2025topological2}. The other is based on $\Delta$ complex by constructing $\Delta$-complexes and classifying spaces, generating persistent homology, and persistent path homology on genome sequences \cite{liu2025topological3}. These methods all had good performance across a variety of tasks, including the phylogenetic analysis of SARS-CoV-2 variants, prediction of protein-nucleic
acid binding affinities, and biological clustering. Suwayyid et al. introduced commutative algebra k-mer learning for genomic sequences analyzing, which bridges between commutative algebra, algebraic topology, combinatorics, and ML to establish a new mathematical paradigm for comparative genomic analysis \cite{suwayyid2025cakl}. 

Identifying novel and functional RNA structures remains a significant challenge in RNA motif design and is critical for advancing RNA-based therapeutics. A computational, topology-based approach combined with unsupervised machine-learning algorithms was used to estimate the size and content of the database of RNA-like graph topologies \cite{wang2025large}. This work provides valuable insights into the scope of the RNA motif universe and informs RNA design strategies, offering a new framework for predicting RNA graph topologies and guiding the discovery of novel RNA motifs—potentially enabling the development of antiviral therapeutics through subgraph assembly.

\subsection{Single cell data analysis}
Single cell data are high-dimensional and sparse, reflecting cellular heterogeneity, rare populations, and dynamic transcriptional states at single cell resolution. TDA offers a powerful framework for single cell transcriptomics by capturing high-dimensional, nonlinear structures beyond traditional clustering \cite{huynh2024topological}. Unlike linear methods, TDA preserves global and local topological features, enabling robust identification of rare cell populations, continuous differentiation trajectories, and hidden functional relationships within complex cellular heterogeneity. For instance, Hodge decomposition method has been developed and successfully applied to RNA velocity field, which captures the cell dynamic information in the biological processes \cite{su2024hodge,su2024topology,su2024hodge2}

Currently, a series of topological-enhanced approaches were proposed to enhance existing approaches in single cell data analysis. For instance, persistent Laplacian-enhanced PCA was proposed to tackle multiscale and multiclass heterogeneity problems in single cell data \cite{cottrell2023knearestneighbors,cottrell2023plpca}. A k-Nearest-Neighbor (kNN) persistent Laplacian (PL) technique was also introduced to improve upon the traditional persistent homology. Unlike traditional persistent homology which filtrates by varying a distance threshold, kNN-PL filtrates by varying the number of neighbors in a kNN network at each step. These techniques were successfully applied on micro-array and single-cell RNA sequencing data \cite{cottrell2023knearestneighbors,cottrell2023plpca}. Thereafter, methods such as topological nonnegative matrix factorization (TNMF) and robust topological nonnegative matrix factorizatiom (rTNMF) were also introduced and outperform all other NMF-based methods in single-cell RNA sequence analysis \cite{hozumi2024analyzing}. Both TNMF and rTNMF are persistent Laplacian regularized NMF methods which can better capture multiscale geometric information and improve performance compared to other NMF-based methods.

Recently, Cottrell and Wei have proposed multiscale cell-cell interactive spatial transcriptomics analysis. This approach integrates the advantages of an ensemble of multiscale topological representations of cell-cell interactions in the gene expression space with those of advanced spatial deep learning techniques \cite{cottrell2025multiscale}.

\subsection{Genetic and genomic network analysis}

Genetic and genomic network analysis explores interactions among genes, regulatory elements, and molecular pathways to reveal system-level biological mechanisms. Traditional methods often struggle with high-dimensional noise, nonlinear dependencies, and incomplete representation of complex biological interactions. TDA-based approaches provide advantages by characterizing high-dimensional, nonlinear, and multiscale structural patterns, enabling robust identification of hidden modules and network dynamics beyond conventional graph-based methods\cite{rabadan2019topological}. 

Recently, Ramos et al. presented a novel method using persistent homology to analyze the role of driver genes in higher-order structures within Cancer Consensus Networks derived from main cellular pathways, which provided an approach to distinguish drivers and cancer-associated genes from passenger genes \cite{ramos2025identifying}. Masoomy et al. applied persistent homology to cancer gene networks, revealing distinct topological deviations, with loops dominating cancer cells and voids prevalent in healthy cells, highlighting higher-order structural differences \cite{masoomy2021topological}. Duman and Pirim employed persistent homology on Arabidopsis weighted gene coexpression networks from microarray data, using bottleneck distances and clustering to distinguish stress responses, demonstrating efficient detection of shared topological features \cite{duman2018gene}. Additionally, Platt et al. used persistent homology to identify phenotypes that tended to be dominated by metabolic syndrome descriptions \cite{platt2016characterizing}.

\section{Outlook}

TDA holds great promise to revolutionize the domain of molecular sciences. The past few decades have allowed TDA to evolve and transformed the domains of macromolecules and interactions, drug design and discovery,  and materials science. To truly tap into this potential and expedite the transition of TDA to other areas of molecular sciences, it is crucial for the research community to maintain this interdisciplinary approach anchored in innovation, collaboration, and the principles of open science. This approach calls for the amalgamation of various fields, uniting experts in mathematics, biology, chemistry, materials, physics, computer science, and machine learning to cultivate a fertile environment for cross-disciplinary interaction. By promoting open communication and the sharing of ideas, TDA is poised to evolve swiftly, overcoming existing limitations and igniting pioneering breakthroughs.

The achievements of TDA in molecular sciences highlighted a few areas in which TDA can be further improved to benefit  molecular sciences. The recently proposed interaction homotopy and interaction homology can be applied in protein-ligand interactions, protein-protein interactions, drug-target interactions and antibody-antigen interactions\cite{liu2023interaction}. The mathematical construction of interaction homology would allow researchers to have a better understanding of interactions. 

Although the topological perturbation analysis (TPA) have been successful in detecting important nodes within various complex networks \cite{chen2023path}, TPA should be further explored in order to design the localized version of TPA, which could enable precise detection of functional modules within complex biological networks. 

Further, the applications of persistent sheaf analysis\cite{wei2025persistent3} and weighted persistent homology should be further explored. For example, these concepts have huge potential for the TDA community to design new hypergraph embeddings that can assign weights or emphasize on molecular functional groups such as benzenes, esters, thiols, amines etc.

Furthermore, differential topology \cite{su2024persistent} and geometric topology \cite{shen2024evolutionary} are expected to play a more important role in molecular sciences. The development of effective computational tools for these approaches is needed. 

Finally, initiated by Cang and Wei in 2017\cite{cang2017topologynet}, topological deep learning (TDL) has become a new frontier in rational learning. There is great potential to further explore this paradigm in molecular sciences, particularly through large language models (LLMs), foundation models, artificial general intelligence (AGI), model context protocol (MCP), and other advanced AI platforms.

\section*{Acknowledgments}

This work was supported in part by National Science Foundation grants DMS2052983  and IIS-1900473, and MSU Research Foundation.  

\bibliography{weig_v2}

\begin{thebibliography}{100}

\bibitem{abramson2024accurate}
J.~Abramson, J.~Adler, J.~Dunger, R.~Evans, T.~Green, A.~Pritzel, O.~Ronneberger, L.~Willmore, A.~J. Ballard, J.~Bambrick, et~al.
\newblock Accurate structure prediction of biomolecular interactions with alphafold 3.
\newblock {\em Nature}, 630(8016):493--500, 2024.

\bibitem{adams2017persistence}
H.~Adams, T.~Emerson, M.~Kirby, R.~Neville, C.~Peterson, P.~Shipman, S.~Chepushtanova, E.~Hanson, F.~Motta, and L.~Ziegelmeier.
\newblock Persistence images: A stable vector representation of persistent homology.
\newblock {\em Journal of Machine Learning Research}, 18:1--35, 2017.

\bibitem{ameneyro2024quantum}
B.~Ameneyro, V.~Maroulas, and G.~Siopsis.
\newblock Quantum persistent homology.
\newblock {\em Journal of Applied and Computational Topology}, (7):1961--1980, 2024.

\bibitem{anand2020weighted}
D.~V. Anand, Z.~Meng, K.~Xia, and Y.~Mu.
\newblock Weighted persistent homology for osmolyte molecular aggregation and hydrogen-bonding network analysis.
\newblock {\em Scientific reports}, 10(1):9685, 2020.

\bibitem{anand2022topological}
D.~V. Anand, Q.~Xu, J.~Wee, K.~Xia, and T.~C. Sum.
\newblock Topological feature engineering for machine learning based halide perovskite materials design.
\newblock {\em npj Computational Materials}, 8(1):203, 2022.

\bibitem{arango2023topological}
A.~S. Arango, H.~Park, and E.~Tajkhorshid.
\newblock Topological learning approach to characterize lipids.
\newblock {\em Biophysical journal}, 122(3):22a, 2023.

\bibitem{arango2024topological}
A.~S. Arango, H.~Park, and E.~Tajkhorshid.
\newblock Topological learning approach to characterizing biological membranes.
\newblock {\em Journal of Chemical Information and Modeling}, 64(13):5242--5252, 2024.

\bibitem{arsuaga2012topological}
J.~Arsuaga, N.~A. Baas, D.~DeWoskin, H.~Mizuno, A.~Pankov, and C.~Park.
\newblock Topological analysis of gene expression arrays identifies high risk molecular subtypes in breast cancer.
\newblock {\em Applicable Algebra in Engineering, Communication and Computing}, 23:3--15, 2012.

\bibitem{bi2023multiscale}
J.~Bi, J.~Wee, X.~Liu, C.~Qu, G.~Wang, and K.~Xia.
\newblock Multiscale {Topological Indices for the Quantitative Prediction of {SARS-CoV}-2 Binding Affinity Change upon M}utations.
\newblock {\em Journal of Chemical Information and Modeling}, 63(13):4216--4227, 2023.

\bibitem{bilsky2025understanding}
J.~Bilsky and A.~E. Clark.
\newblock Understanding the shape of chemistry data—applications with persistent homology.
\newblock {\em The Journal of Chemical Physics}, 163(9), 2025.

\bibitem{bowman2008structural}
G.~R. Bowman, X.~Huang, Y.~Yao, J.~Sun, G.~Carlsson, L.~J. Guibas, and V.~S. Pande.
\newblock Structural insight into rna hairpin folding intermediates.
\newblock {\em Journal of the American Chemical Society}, 130(30):9676--9678, 2008.

\bibitem{bramer2020atom}
D.~Bramer and G.-W. Wei.
\newblock Atom-specific persistent homology and its application to protein flexibility analysis.
\newblock {\em Computational and mathematical biophysics}, 8(1):1--35, 2020.

\bibitem{bubenik2017persistence}
P.~Bubenik and P.~D{\l}otko.
\newblock A persistence landscapes toolbox for topological statistics.
\newblock {\em Journal of Symbolic Computation}, 78:91--114, 2017.

\bibitem{cang2018representability}
Z.~Cang, L.~Mu, and G.-W. Wei.
\newblock Representability of algebraic topology for biomolecules in machine learning based scoring and virtual screening.
\newblock {\em PLoS computational biology}, 14(1):e1005929, 2018.

\bibitem{cang2015topological}
Z.~Cang, L.~Mu, K.~Wu, K.~Opron, K.~Xia, and G.-W. Wei.
\newblock A topological approach for protein classification.
\newblock {\em Computational and Mathematical Biophysics}, 3(1), 2015.

\bibitem{cang2020evolutionary}
Z.~Cang, E.~Munch, and G.-W. Wei.
\newblock Evolutionary homology on coupled dynamical systems with applications to protein flexibility analysis.
\newblock {\em Journal of applied and computational topology}, 4:481--507, 2020.

\bibitem{cang2017analysis}
Z.~Cang and G.-W. Wei.
\newblock Analysis and prediction of protein folding energy changes upon mutation by element specific persistent homology.
\newblock {\em Bioinformatics}, 33(22):3549--3557, 2017.

\bibitem{cang2017topological}
Z.~Cang and G.-W. Wei.
\newblock Topological fingerprints reveal protein-ligand binding mechanism.
\newblock {\em arXiv preprint arXiv:1703.10982}, 2017.

\bibitem{cang2017topologynet}
Z.~Cang and G.-W. Wei.
\newblock Topology{N}et: {T}opology based deep convolutional and multi-task neural networks for biomolecular property predictions.
\newblock {\em PLoS computational biology}, 13(7):e1005690, 2017.

\bibitem{cang2018integration}
Z.~Cang and G.-W. Wei.
\newblock Integration of element specific persistent homology and machine learning for protein-ligand binding affinity prediction.
\newblock {\em International journal for numerical methods in biomedical engineering}, 34(2):e2914, 2018.

\bibitem{cang2020persistent}
Z.~Cang and G.-W. Wei.
\newblock Persistent cohomology for data with multicomponent heterogeneous information.
\newblock {\em SIAM journal on mathematics of data science}, 2(2):396--418, 2020.

\bibitem{carlsson2009topology}
G.~Carlsson.
\newblock Topology and data.
\newblock {\em Bulletin of the American Mathematical Society}, 46(2):255--308, 2009.

\bibitem{chan2013topology}
J.~M. Chan, G.~Carlsson, and R.~Rabadan.
\newblock Topology of viral evolution.
\newblock {\em Proceedings of the National Academy of Sciences}, 110(46):18566--18571, 2013.

\bibitem{chen2025category}
D.~Chen, C.-L. Chen, and G.-W. Wei.
\newblock Category-specific topological learning of metal--organic frameworks.
\newblock {\em Journal of Materials Chemistry A}, 13(13):9292--9303, 2025.

\bibitem{chen2021algebraic}
D.~Chen, K.~Gao, D.~D. Nguyen, X.~Chen, Y.~Jiang, G.-W. Wei, and F.~Pan.
\newblock Algebraic graph-assisted bidirectional transformers for molecular property prediction.
\newblock {\em Nature communications}, 12(1):3521, 2021.

\bibitem{chen2024drug}
D.~Chen, G.~Liu, H.~Du, J.~Wee, R.~Wang, J.~Chen, J.~Shen, and G.-W. Wei.
\newblock Drug resistance revealed by in silico deep mutational scanning and mutation tracker.
\newblock {\em arXiv preprint arXiv:2403.02603}, 2024.

\bibitem{chen2024multiscale}
D.~Chen, J.~Liu, and G.-W. Wei.
\newblock Multiscale topology-enabled structure-to-sequence transformer for protein--ligand interaction predictions.
\newblock {\em Nature Machine Intelligence}, 6(7):799--810, 2024.

\bibitem{chen2024topo}
D.~Chen, J.~Liu, and G.-W. Wei.
\newblock {TopoFormer: Multiscale Topology-enabled Structure-to-Sequence Transformer for Protein-Ligand Interaction Predictions}.
\newblock 2024.

\bibitem{chen2023persistent}
D.~Chen, J.~Liu, J.~Wu, and G.-W. Wei.
\newblock Persistent hyperdigraph homology and persistent hyperdigraph {L}aplacians.
\newblock {\em Foundations of data science}, 5(4):558, 2023.

\bibitem{chen2023path}
D.~Chen, J.~Liu, J.~Wu, G.-W. Wei, F.~Pan, and S.-T. Yau.
\newblock Path topology in molecular and materials sciences.
\newblock {\em The Journal of Physical Chemistry Letters}, 14(4):954--964, 2023.

\bibitem{chen2025superionic}
D.~Chen, B.~Wang, S.~Li, W.~Zhang, K.~Yang, Y.~Song, G.-W. Wei, and F.~Pan.
\newblock Superionic ionic conductor discovery via multiscale topological learning.
\newblock {\em Journal of the American Chemical Society}, 147(24):20888, 2025.

\bibitem{chen2025enhancing}
D.~Chen, R.~Wang, G.-W. Wei, and F.~Pan.
\newblock Enhancing energy predictions in multi-atom systems with multiscale topological learning.
\newblock {\em J. Mater. Chem. A}, 13:21555--21563, 2025.

\bibitem{chen2020persistent}
D.~Chen, M.-Z. Zhang, H.-B. Chen, Z.-W. Xie, G.-W. Wei, and F.~Pan.
\newblock Persistent homology for the quantitative analysis of the structure and stability of carboranes.
\newblock {\em Chinese J. Struc. Chem}, 39(6):999--1008, 2020.

\bibitem{chen2021review}
J.~Chen, K.~Gao, R.~Wang, D.~D. Nguyen, and G.-W. Wei.
\newblock Review of covid-19 antibody therapies.
\newblock {\em Annual review of biophysics}, 50:1--30, 2021.

\bibitem{chen2021prediction}
J.~Chen, K.~Gao, R.~Wang, and G.-W. Wei.
\newblock Prediction and mitigation of mutation threats to covid-19 vaccines and antibody therapies.
\newblock {\em Chemical science}, 12(20):6929--6948, 2021.

\bibitem{chen2021revealing}
J.~Chen, K.~Gao, R.~Wang, and G.-W. Wei.
\newblock Revealing the threat of emerging {SARS-CoV}-2 mutations to antibody therapies.
\newblock {\em Journal of molecular biology}, 433(18):167155, 2021.

\bibitem{chen2022persistent}
J.~Chen, Y.~Qiu, R.~Wang, and G.-W. Wei.
\newblock Persistent {L}aplacian projected {O}micron {BA. 4 and BA. 5} to become new dominating variants.
\newblock {\em Computers in Biology and Medicine}, 151:106262, 2022.

\bibitem{chen2022emerging}
J.~Chen, R.~Wang, Y.~Hozumi, G.~Liu, Y.~Qiu, X.~Wei, and G.-W. Wei.
\newblock Emerging dominant {SARS-CoV}-2 variants.
\newblock {\em Journal of Chemical Information and Modeling}, 63(1):335--342, 2022.

\bibitem{chen2020mutations}
J.~Chen, R.~Wang, M.~Wang, and G.-W. Wei.
\newblock Mutations strengthened {SARS-CoV}-2 infectivity.
\newblock {\em Journal of molecular biology}, 432(19):5212--5226, 2020.

\bibitem{chen2021sars}
J.~Chen, R.~Wang, and G.-W. Wei.
\newblock {SARS-CoV}-2 becoming more infectious as revealed by algebraic topology and deep learning.
\newblock {\em Communications in information and systems}, 21(1):31, 2021.

\bibitem{chen2022mathematical}
J.~Chen and G.-W. Wei.
\newblock Mathematical artificial intelligence design of mutation-proof {COVID}-19 monoclonal antibodies.
\newblock {\em Communications in information and systems}, 22(3):339, 2022.

\bibitem{chen2022omicron}
J.~Chen and G.-W. Wei.
\newblock Omicron ba. 2 (b. 1.1. 529.2): high potential for becoming the next dominant variant.
\newblock {\em The journal of physical chemistry letters}, 13(17):3840--3849, 2022.

\bibitem{chen2023topological}
J.~Chen, D.~R. Woldring, F.~Huang, X.~Huang, and G.-W. Wei.
\newblock Topological deep learning based deep mutational scanning.
\newblock {\em Computers in Biology and Medicine}, 164:107258, 2023.

\bibitem{chen2021evolutionary}
J.~Chen, R.~Zhao, Y.~Tong, and G.-W. Wei.
\newblock Evolutionary de {Rham-H}odge method.
\newblock {\em Discrete and continuous dynamical systems. Series B}, 26(7):3785, 2021.

\bibitem{chen2020topology}
X.~Chen, D.~Chen, M.~Weng, Y.~Jiang, G.-W. Wei, and F.~Pan.
\newblock Topology-based machine learning strategy for cluster structure prediction.
\newblock {\em The journal of physical chemistry letters}, 11(11):4392--4401, 2020.

\bibitem{christensen_medium-range_2023}
R.~Christensen, Y.~Bokor~Bleile, S.~S. Sørensen, C.~A.~N. Biscio, L.~Fajstrup, and M.~M. Smedskjaer.
\newblock Medium-range order structure controls thermal stability of pores in zeolitic imidazolate frameworks.
\newblock {\em The Journal of Physical Chemistry Letters}, 14(33):7469--7476.

\bibitem{cottrell2023knearestneighbors}
S.~Cottrell, Y.~Hozumi, and G.-W. Wei.
\newblock K-{Nearest-Neighbors Induced Topological PCA for scRNA Sequence Data Analysis}.
\newblock {\em Computer in biology and medicine}, 175:108497, 2023.

\bibitem{cottrell2023plpca}
S.~Cottrell, R.~Wang, and G.-W. Wei.
\newblock P{LPCA: Persistent Laplacian Enhanced-PCA for Microarray Data Analysis}.
\newblock {\em Journal of chemical information and modeling}, 64(7):2405--2420, 2023.

\bibitem{cottrell2025multiscale}
S.~Cottrell and G.-W. Wei.
\newblock Multiscale cell-cell interactive spatial transcriptomics analysis.
\newblock {\em Research Square}, pages rs--3, 2025.

\bibitem{dai2025topological}
X.~Dai, R.~Wang, and Y.~Zhang.
\newblock Topological deep learning for enhancing peptide-protein complex prediction.
\newblock {\em Communications Chemistry}, page Accepted, 2025.

\bibitem{dong2025exploring}
Y.~Dong, M.~Xu, and L.~Tang.
\newblock Exploring multi-modal representations based on an improved multi-objective optimization algorithm for molecular property prediction.
\newblock In {\em Proceedings of the Genetic and Evolutionary Computation Conference Companion}, pages 711--714, 2025.

\bibitem{dou2023machine}
B.~Dou, Z.~Zhu, E.~Merkurjev, L.~Ke, L.~Chen, J.~Jiang, Y.~Zhu, J.~Liu, B.~Zhang, and G.-W. Wei.
\newblock Machine learning methods for small data challenges in molecular science.
\newblock {\em Chemical Reviews}, 123(13):8736--8780, 2023.

\bibitem{du2024multiscale}
H.~Du, G.-W. Wei, and T.~Hou.
\newblock Multiscale topology in interactomic network: from transcriptome to antiaddiction drug repurposing.
\newblock {\em Briefings in Bioinformatics}, 25(2):bbae054, 2024.

\bibitem{duman2018gene}
A.~N. Duman and H.~Pirim.
\newblock Gene coexpression network comparison via persistent homology.
\newblock {\em International journal of genomics}, 2018(1):7329576, 2018.

\bibitem{edelsbrunner2008persistent}
H.~Edelsbrunner, J.~Harer, et~al.
\newblock Persistent homology-a survey.
\newblock {\em Contemporary mathematics}, 453(26):257--282, 2008.

\bibitem{ehiro2024descriptor}
T.~Ehiro.
\newblock Descriptor generation from morgan fingerprint using persistent homology.
\newblock {\em SAR and QSAR in Environmental Research}, 35(1):31--51, 2024.

\bibitem{fang2025leveraging}
Z.~Fang and Q.~Yan.
\newblock Leveraging persistent homology features for accurate defect formation energy predictions via graph neural networks.
\newblock {\em Chemistry of materials}, 37(4):1531--1540, 2025.

\bibitem{feng2025mayer}
H.~Feng, L.~Shen, J.~Liu, and G.-W. Wei.
\newblock Mayer-homology learning prediction of protein-ligand binding affinities.
\newblock {\em Journal of Computational Biophysics and Chemistry}, 24(2):253--266, 2025.

\bibitem{feng2025caml}
H.~Feng, F.~Suwayyid, M.~Zia, J.~Wee, Y.~Hozumi, C.-L. Chen, and G.-W. Wei.
\newblock Caml: Commutative algebra machine learning-a case study on protein-ligand binding affinity prediction.
\newblock {\em Journal of Chemical Information and Modeling}, 65(13):6732--6743, 2025.

\bibitem{feng2023virtual}
H.~Feng and G.-W. Wei.
\newblock Virtual screening of {DrugB}ank database for herg blockers using topological laplacian-assisted {AI} models.
\newblock {\em Computers in biology and medicine}, 153:106491, 2023.

\bibitem{gameiro2015topological}
M.~Gameiro, Y.~Hiraoka, S.~Izumi, M.~Kramar, K.~Mischaikow, and V.~Nanda.
\newblock A topological measurement of protein compressibility.
\newblock {\em Japan Journal of Industrial and Applied Mathematics}, 32:1--17, 2015.

\bibitem{gao2022methodology}
K.~Gao, R.~Wang, J.~Chen, L.~Cheng, J.~Frishcosy, Y.~Huzumi, Y.~Qiu, T.~Schluckbier, X.~Wei, and G.-W. Wei.
\newblock Methodology-centered review of molecular modeling, simulation, and prediction of {SARS-CoV}-2.
\newblock {\em Chemical Reviews}, 122(13):11287--11368, 2022.

\bibitem{gao2021persistent}
Y.~Gao, F.~Lei, and S.~X. Li.
\newblock Persistent homology and application on residues 1 to 28 of amyloid beta peptide.
\newblock {\em Proteins: Structure, Function, and Bioinformatics}, 89(4):409--415, 2021.

\bibitem{ghrist2008barcodes}
R.~Ghrist.
\newblock Barcodes: the persistent topology of data.
\newblock {\em Bulletin of the American Mathematical Society}, 45(1):61--75, 2008.

\bibitem{grottel2011topological}
S.~Grottel, C.~A. Dietrich, J.~L. Comba, and T.~Ertl.
\newblock Topological extraction and tracking of defects in crystal structures.
\newblock In {\em Topological Methods in Data Analysis and Visualization: Theory, Algorithms, and Applications}, pages 167--178. Springer, 2010.

\bibitem{han2025topoqa}
B.~Han, Y.~Zhang, L.~Li, X.~Gong, and K.~Xia.
\newblock Topoqa: a topological deep learning-based approach for protein complex structure interface quality assessment.
\newblock {\em Briefings in Bioinformatics}, 26(2):bbaf083, 2025.

\bibitem{harvey2014collaborative}
W.~Harvey, I.-H. Park, O.~R{\"u}bel, V.~Pascucci, P.-T. Bremer, C.~Li, and Y.~Wang.
\newblock A collaborative visual analytics suite for protein folding research.
\newblock {\em Journal of Molecular Graphics and Modelling}, 53:59--71, 2014.

\bibitem{hayes2025persistent}
N.~Hayes, X.~Wei, H.~Feng, E.~Merkurjev, and G.-W. Wei.
\newblock Persistent sheaf laplacian analysis of protein flexibility.
\newblock {\em The Journal of Physical Chemistry B}, 129(17):4169--4178, 2025.

\bibitem{higashi2025feature}
Y.~Higashi, K.~Okuwaki, Y.~Mochizuki, T.~Fujigaya, and K.~Kato.
\newblock Feature vectorization of microphase-separated structures in polymeric materials using dissipative particle dynamics and persistent homology for machine learning applications.
\newblock {\em Digital Discovery}, 4(5):1339--1351, 2025.

\bibitem{hiraoka2016hierarchical}
Y.~Hiraoka, T.~Nakamura, A.~Hirata, E.~G. Escolar, K.~Matsue, and Y.~Nishiura.
\newblock Hierarchical structures of amorphous solids characterized by persistent homology.
\newblock {\em Proceedings of the National Academy of Sciences}, 113(26):7035--7040, 2016.

\bibitem{hong2019medium}
S.~Hong and D.~Kim.
\newblock Medium-range order in amorphous ices revealed by persistent homology.
\newblock {\em Journal of Physics: Condensed Matter}, 31(45):455403, 2019.

\bibitem{hozumi2024analyzing}
Y.~Hozumi and G.-W. Wei.
\newblock Analyzing single cell rna sequencing with topological nonnegative matrix factorization.
\newblock {\em Journal of Computational and Applied Mathematics}, 445:115842, 2024.

\bibitem{hozumi2024revealing}
Y.~Hozumi and G.-W. Wei.
\newblock Revealing the shape of genome space via k-mer topology.
\newblock {\em arXiv preprint arXiv:2412.20202}, 2024.

\bibitem{huynh2024topological}
T.~Huynh and Z.~Cang.
\newblock Topological and geometric analysis of cell states in single-cell transcriptomic data.
\newblock {\em Briefings in Bioinformatics}, 25(3):bbae176, 2024.

\bibitem{ichinomiya2022topological}
T.~Ichinomiya.
\newblock Topological data analysis gives two folding paths in hp35 (nle-nle), double mutant of villin headpiece subdomain.
\newblock {\em Scientific Reports}, 12(1):2719, 2022.

\bibitem{ichinomiya2020protein}
T.~Ichinomiya, I.~Obayashi, and Y.~Hiraoka.
\newblock Protein-folding analysis using features obtained by persistent homology.
\newblock {\em Biophysical Journal}, 118(12):2926--2937, 2020.

\bibitem{jiang2022molecular}
P.~Jiang, Y.~Chi, X.-S. Li, Z.~Meng, X.~Liu, X.-S. Hua, and K.~Xia.
\newblock Molecular persistent spectral image ({Mol-PSI}) representation for machine learning models in drug design.
\newblock {\em Briefings in Bioinformatics}, 23(1):bbab527, 2022.

\bibitem{jiang2021topological}
Y.~Jiang, D.~Chen, X.~Chen, T.~Li, G.-W. Wei, and F.~Pan.
\newblock Topological representations of crystalline compounds for the machine-learning prediction of materials properties.
\newblock {\em npj computational materials}, 7(1):28, 2021.

\bibitem{jones2025petls}
B.~Jones and G.-W. Wei.
\newblock {PETLS: PErsistent Topological Laplacian Software}.
\newblock {\em arXiv preprint arXiv:2508.11560}, 2025.

\bibitem{keller2018phos}
B.~Keller, M.~Lesnick, and T.~L. Willke.
\newblock P{HoS}: persistent homology for virtual screening.
\newblock 2018.

\bibitem{koseki2023topological}
J.~Koseki, S.~Hayashi, Y.~Kojima, H.~Hirose, and T.~Shimamura.
\newblock Topological data analysis of protein structure and inter/intra-molecular interaction changes attributable to amino acid mutations.
\newblock {\em Computational and Structural Biotechnology Journal}, 21:2950--2959, 2023.

\bibitem{kovacev2016using}
V.~Kovacev-Nikolic, P.~Bubenik, D.~Nikoli{\'c}, and G.~Heo.
\newblock Using persistent homology and dynamical distances to analyze protein binding.
\newblock {\em Statistical applications in genetics and molecular biology}, 15(1):19--38, 2016.

\bibitem{krishnapriyan2020topological}
A.~S. Krishnapriyan, M.~Haranczyk, and D.~Morozov.
\newblock Topological descriptors help predict guest adsorption in nanoporous materials.
\newblock {\em The Journal of Physical Chemistry C}, 124(17):9360--9368, 2020.

\bibitem{krishnapriyan2021machine}
A.~S. Krishnapriyan, J.~Montoya, M.~Haranczyk, J.~Hummelsh{\o}j, and D.~Morozov.
\newblock Machine learning with persistent homology and chemical word embeddings improves prediction accuracy and interpretability in metal-organic frameworks.
\newblock {\em Scientific reports}, 11(1):8888, 2021.

\bibitem{landuzzi2020persistence}
F.~Landuzzi, T.~Nakamura, D.~Michieletto, and T.~Sakaue.
\newblock Persistence homology of entangled rings.
\newblock {\em Physical Review Research}, 2(3):033529, 2020.

\bibitem{liang1998anatomy}
J.~Liang, C.~Woodward, and H.~Edelsbrunner.
\newblock Anatomy of protein pockets and cavities: measurement of binding site geometry and implications for ligand design.
\newblock {\em Protein science}, 7(9):1884--1897, 1998.

\bibitem{liu2023interaction}
J.~Liu, D.~Chen, and G.-W. Wei.
\newblock Interaction homotopy and interaction homology.
\newblock {\em arXiv preprint arXiv:2311.16322}, 2023.

\bibitem{liu2025persistent}
J.~Liu, D.~Chen, and G.-W. Wei.
\newblock Persistent interaction topology in data analysis.
\newblock {\em Foundations of Data Science}, page doi: 10.3934/fods.2025011, 2025.

\bibitem{liu2023algebraic}
J.~Liu, J.~Li, and J.~Wu.
\newblock The algebraic stability for persistent laplacians.
\newblock {\em Homology, Homotopy \& Applications}, 26(2), 2024.

\bibitem{liu2025topological3}
J.~Liu, L.~Shen, D.~Chen, and G.-W. Wei.
\newblock Topological sequence analysis of genomes: Delta complex approaches.
\newblock {\em arXiv preprint arXiv:2507.05452}, 2025.

\bibitem{liu2024persistent}
J.~Liu, L.~Shen, and G.-W. Wei.
\newblock Persistent khovanov homology of tangles.
\newblock {\em arXiv preprint arXiv:2409.18312}, 2024.

\bibitem{liu2025topological2}
J.~Liu, L.~Shen, M.~Zia, and G.-W. Wei.
\newblock Topological sequence analysis of genomes: Category approaches.
\newblock {\em arXiv preprint arXiv:2507.08043}, 2025.

\bibitem{liu2023persistent}
R.~Liu, X.~Liu, and J.~Wu.
\newblock Persistent path-spectral (pps) based machine learning for protein--ligand binding affinity prediction.
\newblock {\em Journal of Chemical Information and Modeling}, 63(3):1066--1075, 2023.

\bibitem{liu2021persistent}
X.~Liu, H.~Feng, J.~Wu, and K.~Xia.
\newblock Persistent spectral hypergraph based machine learning ({PSH-ML}) for protein-ligand binding affinity prediction.
\newblock {\em Briefings in Bioinformatics}, 22(5):bbab127, 2021.

\bibitem{liu2022dowker}
X.~Liu, H.~Feng, J.~Wu, and K.~Xia.
\newblock Dowker complex based machine learning ({DCML}) models for protein-ligand binding affinity prediction.
\newblock {\em PLoS Computational Biology}, 18(4):e1009943, 2022.

\bibitem{liu2022hom}
X.~Liu, H.~Feng, J.~Wu, and K.~Xia.
\newblock Hom-complex-based machine learning ({HCML}) for the prediction of protein--protein binding affinity changes upon mutation.
\newblock {\em Journal of chemical information and modeling}, 62(17):3961--3969, 2022.

\bibitem{liu2025machine}
X.~Liu, X.~Huang, and G.-W. Wei.
\newblock Machine-learning prediction of virus-like particle stoichiometry and stability using persistent topological laplacians.
\newblock {\em arXiv preprint arXiv:2507.21417}, 2025.

\bibitem{liu2025manifold}
X.~Liu, Z.~Su, Y.~Shi, Y.~Tong, G.~Wang, and G.-W. Wei.
\newblock Manifold topological deep learning for biomedical data.
\newblock {\em arXiv preprint arXiv:2503.00175}, 2025.

\bibitem{liu2021hypergraph}
X.~Liu, X.~Wang, J.~Wu, and K.~Xia.
\newblock Hypergraph-based persistent cohomology ({HPC}) for molecular representations in drug design.
\newblock {\em Briefings in Bioinformatics}, 22(5):bbaa411, 2021.

\bibitem{liu2025topological}
X.~Liu, J.~Wee, and G.-W. Wei.
\newblock Topological machine learning for protein-nucleic acid binding affinity changes upon mutation.
\newblock {\em arXiv preprint arXiv:2505.22786}, 2025.

\bibitem{liu2021neighborhood}
X.~Liu and K.~Xia.
\newblock Neighborhood complex based machine learning (ncml) models for drug design.
\newblock In {\em Interpretability of Machine Intelligence in Medical Image Computing, and Topological Data Analysis and Its Applications for Medical Data: 4th International Workshop, iMIMIC 2021, and 1st International Workshop, TDA4MedicalData 2021, Held in Conjunction with MICCAI 2021, Strasbourg, France, September 27, 2021, Proceedings 4}, pages 87--97. Springer, 2021.

\bibitem{liu2022persistent}
X.~Liu and K.~Xia.
\newblock Persistent {T}or-algebra based stacking ensemble learning ({PTA-SEL}s) for protein-protein binding affinity prediction.
\newblock In {\em Topological, Algebraic and Geometric Learning Workshops 2022}, pages 237--247. PMLR, 2022.

\bibitem{long2023predicting}
Y.~Long and B.~R. Donald.
\newblock Predicting affinity through homology (path): Interpretable binding affinity prediction with persistent homology.
\newblock {\em bioRxiv}, pages 2023--11, 2023.

\bibitem{masoomy2021topological}
H.~Masoomy, B.~Askari, S.~Tajik, A.~K. Rizi, and G.~R. Jafari.
\newblock Topological analysis of interaction patterns in cancer-specific gene regulatory network: persistent homology approach.
\newblock {\em Scientific Reports}, 11(1):16414, 2021.

\bibitem{membrillo2022tracking}
I.~Membrillo~Solis, T.~Orlova, K.~Bednarska, P.~Lesiak, T.~R. Woli{\'n}ski, G.~D’Alessandro, J.~Brodzki, and M.~Kaczmarek.
\newblock Tracking the time evolution of soft matter systems via topological structural heterogeneity.
\newblock {\em Communications Materials}, 3(1):1, 2022.

\bibitem{memoli2022persistent}
F.~M{\'e}moli, Z.~Wan, and Y.~Wang.
\newblock Persistent laplacians: Properties, algorithms and implications.
\newblock {\em SIAM Journal on Mathematics of Data Science}, 4(2):858--884, 2022.

\bibitem{meng2020weighted}
Z.~Meng, D.~V. Anand, Y.~Lu, J.~Wu, and K.~Xia.
\newblock Weighted persistent homology for biomolecular data analysis.
\newblock {\em Scientific reports}, 10(1):2079, 2020.

\bibitem{meng2021persistent}
Z.~Meng and K.~Xia.
\newblock Persistent spectral--based machine learning ({PerSpect ML}) for protein-ligand binding affinity prediction.
\newblock {\em Science advances}, 7(19):eabc5329, 2021.

\bibitem{minamitani2023persistent}
E.~Minamitani, I.~Obayashi, K.~Shimizu, and S.~Watanabe.
\newblock Persistent homology-based descriptor for machine-learning potential of amorphous structures.
\newblock {\em The Journal of Chemical Physics}, 159(8), 2023.

\bibitem{mirebrahimi2019persistent}
H.~Mirebrahimi and A.~Babaee.
\newblock Persistent homology for prediction of protein folding.
\newblock In {\em 1st Annual National Conference on Biomathematics}, page~88, 2019.

\bibitem{mirth2021representations}
J.~Mirth, Y.~Zhai, J.~Bush, E.~G. Alvarado, H.~Jordan, M.~Heim, B.~Krishnamoorthy, M.~Pflaum, A.~Clark, H.~Adams, et~al.
\newblock Representations of energy landscapes by sublevelset persistent homology: an example with n-alkanes.
\newblock {\em The Journal of Chemical Physics}, 154(11), 2021.

\bibitem{morley2021persistent}
D.~O. Morley, P.~S. Salmon, and M.~Wilson.
\newblock Persistent homology in two-dimensional atomic networks.
\newblock {\em The Journal of Chemical Physics}, 154(12):124109, 2021.

\bibitem{nakamura2015persistent}
T.~Nakamura, Y.~Hiraoka, A.~Hirata, E.~G. Escolar, and Y.~Nishiura.
\newblock Persistent homology and many-body atomic structure for medium-range order in the glass.
\newblock {\em Nanotechnology}, 26(30):304001, 2015.

\bibitem{nguyen2020review}
D.~D. Nguyen, Z.~Cang, and G.-W. Wei.
\newblock A review of mathematical representations of biomolecular data.
\newblock {\em Physical Chemistry Chemical Physics}, 22(8):4343--4367, 2020.

\bibitem{nguyen2019mathematical}
D.~D. Nguyen, Z.~Cang, K.~Wu, M.~Wang, Y.~Cao, and G.-W. Wei.
\newblock Mathematical deep learning for pose and binding affinity prediction and ranking in d3r grand challenges.
\newblock {\em Journal of computer-aided molecular design}, 33:71--82, 2019.

\bibitem{nguyen2020unveiling}
D.~D. Nguyen, K.~Gao, J.~Chen, R.~Wang, and G.-W. Wei.
\newblock Unveiling the molecular mechanism of {SARS-CoV}-2 main protease inhibition from 137 crystal structures using algebraic topology and deep learning.
\newblock {\em Chemical science}, 11(44):12036--12046, 2020.

\bibitem{nguyen2020mathdl}
D.~D. Nguyen, K.~Gao, M.~Wang, and G.-W. Wei.
\newblock M{athDL}: mathematical deep learning for {D3R Grand C}hallenge 4.
\newblock {\em Journal of computer-aided molecular design}, 34:131--147, 2020.

\bibitem{nguyen2019agl}
D.~D. Nguyen and G.-W. Wei.
\newblock A{GL}-score: algebraic graph learning score for protein--ligand binding scoring, ranking, docking, and screening.
\newblock {\em Journal of chemical information and modeling}, 59(7):3291--3304, 2019.

\bibitem{obayashi2022persistent}
I.~Obayashi, T.~Nakamura, and Y.~Hiraoka.
\newblock Persistent homology analysis for materials research and persistent homology software: {HomCloud}.
\newblock {\em journal of the physical society of japan}, 91(9):091013, 2022.

\bibitem{offroy2025toxicity}
M.~Offroy, L.~Duponchel, A.~Razafitianamaharavo, C.~Pagnout, and J.~F. Duval.
\newblock Toxicity assessment of titanium dioxide nanoparticles on microorganisms through topological data analysis of high dimensional single-cell nanomechanical data.
\newblock {\em Talanta}, 286:127482, 2025.

\bibitem{onodera2019understanding}
Y.~Onodera, S.~Kohara, S.~Tahara, A.~Masuno, H.~Inoue, M.~Shiga, A.~Hirata, K.~Tsuchiya, Y.~Hiraoka, I.~Obayashi, et~al.
\newblock Understanding diffraction patterns of glassy, liquid and amorphous materials via persistent homology analyses.
\newblock {\em Journal of the Ceramic Society of Japan}, 127(12):853--863, 2019.

\bibitem{papamarkou2024position}
T.~Papamarkou, T.~Birdal, M.~Bronstein, G.~Carlsson, J.~Curry, Y.~Gao, M.~Hajij, R.~Kwitt, P.~Lio, P.~Di~Lorenzo, et~al.
\newblock Position: Topological deep learning is the new frontier for relational learning.
\newblock {\em Proceedings of machine learning research}, 235:39529, 2024.

\bibitem{pirashvili2018improved}
M.~Pirashvili, L.~Steinberg, F.~Belchi~Guillamon, M.~Niranjan, J.~G. Frey, and J.~Brodzki.
\newblock Improved understanding of aqueous solubility modeling through topological data analysis.
\newblock {\em Journal of cheminformatics}, 10(1):1--14, 2018.

\bibitem{platt2016characterizing}
D.~E. Platt, S.~Basu, P.~A. Zalloua, and L.~Parida.
\newblock Characterizing redescriptions using persistent homology to isolate genetic pathways contributing to pathogenesis.
\newblock {\em BMC systems biology}, 10(Suppl 1):S10, 2016.

\bibitem{pun2022persistent}
C.~S. Pun, S.~X. Lee, and K.~Xia.
\newblock Persistent-homology-based machine learning: a survey and a comparative study.
\newblock {\em Artificial Intelligence Review}, 55(7):5169--5213, 2022.

\bibitem{pun2020weighted}
C.~S. Pun, B.~Y.~S. Yong, and K.~Xia.
\newblock Weighted-persistent-homology-based machine learning for {RNA} flexibility analysis.
\newblock {\em PloS one}, 15(8):e0237747, 2020.

\bibitem{qiu2023artificial}
Y.~Qiu and G.-W. Wei.
\newblock Artificial intelligence-aided protein engineering: from topological data analysis to deep protein language models.
\newblock {\em Briefings in Bioinformatics}, 24(5):bbad289, 2023.

\bibitem{qiu2023persistent}
Y.~Qiu and G.-W. Wei.
\newblock Persistent spectral theory-guided protein engineering.
\newblock {\em Nature Computational Science}, 3(2):149--163, 2023.

\bibitem{rabadan2019topological}
R.~Rabad{\'a}n and A.~J. Blumberg.
\newblock {\em Topological data analysis for genomics and evolution: topology in biology}.
\newblock Cambridge University Press, 2019.

\bibitem{ramos2025identifying}
R.~H. Ramos, Y.~A. Bardelotte, C.~de~Oliveira Lage~Ferreira, and A.~Simao.
\newblock Identifying key genes in cancer networks using persistent homology.
\newblock {\em Scientific Reports}, 15(1):2751, 2025.

\bibitem{rong2025topological}
X.~Rong, X.~Haotian, W.~Junwei, Z.~Shufei, S.~Mingjie, L.~Jiejie, and Z.~Quan.
\newblock Topological fusion model for molecular property prediction.
\newblock {\em Applied Intelligence}, 55(11):819, 2025.

\bibitem{sasaki2018liquid}
K.~Sasaki, R.~Okajima, and T.~Yamashita.
\newblock Liquid structures characterized by a combination of the persistent homology analysis and molecular dynamics simulation.
\newblock In {\em AIP Conference Proceedings}, volume 2040. AIP Publishing, 2018.

\bibitem{shekhar2024topological}
S.~Shekhar and C.~Chowdhury.
\newblock Topological data analysis enhanced prediction of hydrogen storage in metal--organic frameworks (mofs).
\newblock {\em Materials Advances}, 5(2):820--830, 2024.

\bibitem{shen2024knot}
L.~Shen, H.~Feng, F.~Li, F.~Lei, J.~Wu, and G.-W. Wei.
\newblock Knot data analysis using multiscale gauss link integral.
\newblock {\em Proceedings of the National Academy of Sciences}, 121(42):e2408431121, 2024.

\bibitem{shen2023svsbi}
L.~Shen, H.~Feng, Y.~Qiu, and G.-W. Wei.
\newblock S{VSBI}: sequence-based virtual screening of biomolecular interactions.
\newblock {\em Communications Biology}, 6(1):536, 2023.

\bibitem{shen2024evolutionary}
L.~Shen, J.~Liu, and G.-W. Wei.
\newblock Evolutionary khovanov homology.
\newblock {\em AIMS Mathematics}, 9(9):26139--26165, 2024.

\bibitem{shen2024persistent}
L.~Shen, J.~Liu, and G.-W. Wei.
\newblock Persistent mayer homology and persistent mayer laplacian.
\newblock {\em Foundations of Data Science}, 6(4):584--612, 2024.

\bibitem{shen2025computing}
L.~Shen, J.~Liu, and G.-W. Wei.
\newblock Computing khovanov homology of tangles.
\newblock {\em arXiv preprint arXiv:2508.14398}, 2025.

\bibitem{shen2025khovanov}
L.~Shen, J.~Liu, and G.-W. Wei.
\newblock Khovanov homology of tangles: algorithm and computation.
\newblock {\em arXiv preprint arXiv:2508.14404}, 2025.

\bibitem{shimizu2021higher}
Y.~Shimizu, T.~Kurokawa, H.~Arai, and H.~Washizu.
\newblock Higher-order structure of polymer melt described by persistent homology.
\newblock {\em Scientific reports}, 11(1):2274, 2021.

\bibitem{song2025multi}
R.~Song, F.~Li, J.~Wu, F.~Lei, and G.-W. Wei.
\newblock Multi-scale jones polynomial and persistent jones polynomial for knot data analysis.
\newblock {\em AIMS mathematics}, 10(1):1463, 2025.

\bibitem{sorensen2020revealing}
S.~S. S{\o}rensen, C.~A. Biscio, M.~Bauchy, L.~Fajstrup, and M.~M. Smedskjaer.
\newblock Revealing hidden medium-range order in amorphous materials using topological data analysis.
\newblock {\em Science Advances}, 6(37):eabc2320, 2020.

\bibitem{spannaus2021materials}
A.~Spannaus, K.~J. Law, P.~Luszczek, F.~Nasrin, C.~P. Micucci, P.~K. Liaw, L.~J. Santodonato, D.~J. Keffer, and V.~Maroulas.
\newblock Materials fingerprinting classification.
\newblock {\em Computer Physics Communications}, 266:108019, 2021.

\bibitem{spirandelli2025topological}
I.~Spirandelli, D.~Morozov, A.~Nigmetov, and M.~E. Evans.
\newblock Topological potentials guiding protein self-assembly.
\newblock {\em arXiv preprint arXiv:2508.15321}, 2025.

\bibitem{stellhorn2020structure}
J.~R. Stellhorn, B.~Paulus, S.~Hosokawa, W.-C. Pilgrim, N.~Boudet, N.~Blanc, H.~Ikemoto, S.~Kohara, and Y.~Sutou.
\newblock Structure of amorphous cu 2 gete 3 and the implications for its phase-change properties.
\newblock {\em Physical Review B}, 101(21):214110, 2020.

\bibitem{su2025topological}
Z.~Su, X.~Liu, L.~B. Hamdan, V.~Maroulas, J.~Wu, G.~Carlsson, and G.-W. Wei.
\newblock Topological data analysis and topological deep learning beyond persistent homology-a review.
\newblock {\em arXiv preprint arXiv:2507.19504}, 2025.

\bibitem{su2024hodge}
Z.~Su, Y.~Tong, and G.-W. Wei.
\newblock Hodge decomposition of single-cell rna velocity.
\newblock {\em Journal of chemical information and modeling}, 64(8):3558--3568, 2024.

\bibitem{su2024hodge2}
Z.~Su, Y.~Tong, and G.-W. Wei.
\newblock Hodge decomposition of vector fields in cartesian grids.
\newblock In {\em SIGGRAPH Asia 2024 Conference Papers}, pages 1--10, 2024.

\bibitem{su2024persistent}
Z.~Su, Y.~Tong, and G.-W. Wei.
\newblock Persistent de rham-hodge laplacians in eulerian representation for manifold topological learning.
\newblock {\em AIMS Mathematics}, 9(10):27438--27470, 2024.

\bibitem{su2024topology}
Z.~Su, Y.~Tong, and G.-W. Wei.
\newblock Topology-preserving hodge decomposition in the eulerian representation.
\newblock {\em arXiv preprint arXiv:2408.14356}, 2024.

\bibitem{suwayyid2025cakl}
F.~Suwayyid, Y.~Hozumi, H.~Feng, M.~Zia, J.~Wee, and G.-W. Wei.
\newblock Cakl: Commutative algebra k-mer learning of genomics.
\newblock {\em arXiv preprint arXiv:2508.09406}, 2025.

\bibitem{suwayyid2024persistent}
F.~Suwayyid and G.-W. Wei.
\newblock Persistent dirac of paths on digraphs and hypergraphs.
\newblock {\em Foundations of Data Science}, 6(2):124--153, 2024.

\bibitem{suwayyid2024persistent2}
F.~Suwayyid and G.-W. Wei.
\newblock Persistent mayer dirac.
\newblock {\em Journal of Physics: Complexity}, 5(4):045005, 2024.

\bibitem{tarin2023computer}
A.~Tar{\'\i}n-Pell{\'o}, B.~Suay-Garc{\'\i}a, J.~For{\'e}s-Martos, A.~Falc{\'o}, and M.-T. P{\'e}rez-Gracia.
\newblock Computer-aided drug repurposing to tackle antibiotic resistance based on topological data analysis.
\newblock {\em Computers in Biology and Medicine}, 166:107496, 2023.

\bibitem{tola2024identification}
A.~Tola, S.~Aziz, D.~Zhabilov, D.~Winkler, B.~Coskunuzer, and M.~Candas.
\newblock Identification of molecular compounds targeting bacterial propionate metabolism with topological machine learning.
\newblock {\em bioRxiv}, pages 2024--10, 2024.

\bibitem{townsend2020representation}
J.~Townsend, C.~P. Micucci, J.~H. Hymel, V.~Maroulas, and K.~D. Vogiatzis.
\newblock Representation of molecular structures with persistent homology for machine learning applications in chemistry.
\newblock {\em Nature communications}, 11(1):3230, 2020.

\bibitem{verovvsek2016extended}
S.~K. Verov{\v{s}}ek and A.~Mashaghi.
\newblock Extended topological persistence and contact arrangements in folded linear molecules.
\newblock {\em Frontiers in Applied Mathematics and Statistics}, 2:6, 2016.

\bibitem{vipond2021multiparameter}
O.~Vipond, J.~A. Bull, P.~S. Macklin, U.~Tillmann, C.~W. Pugh, H.~M. Byrne, and H.~A. Harrington.
\newblock Multiparameter persistent homology landscapes identify immune cell spatial patterns in tumors.
\newblock {\em Proceedings of the National Academy of Sciences}, 118(41):e2102166118, 2021.

\bibitem{wang2025structural}
B.~Wang, B.~Feng, L.~Lv, S.~Li, and F.~Pan.
\newblock Structural feature extraction via topological data analysis.
\newblock {\em The Journal of Physical Chemistry Letters}, 16:8056--8067, 2025.

\bibitem{wang2016object}
B.~Wang and G.-W. Wei.
\newblock Object-oriented persistent homology.
\newblock {\em Journal of computational physics}, 305:276--299, 2016.

\bibitem{wang2020topology}
M.~Wang, Z.~Cang, and G.-W. Wei.
\newblock A topology-based network tree for the prediction of protein--protein binding affinity changes following mutation.
\newblock {\em Nature Machine Intelligence}, 2(2):116--123, 2020.

\bibitem{wang2021analysis}
R.~Wang, J.~Chen, K.~Gao, Y.~Hozumi, C.~Yin, and G.-W. Wei.
\newblock Analysis of {SARS-CoV}-2 mutations in the united states suggests presence of four substrains and novel variants.
\newblock {\em Communications biology}, 4(1):228, 2021.

\bibitem{wang2021mechanisms}
R.~Wang, J.~Chen, and G.-W. Wei.
\newblock Mechanisms of {SARS-CoV}-2 evolution revealing vaccine-resistant mutations in europe and america.
\newblock {\em The journal of physical chemistry letters}, 12(49):11850--11857, 2021.

\bibitem{wang2023chatgpt}
R.~Wang, H.~Feng, and G.-W. Wei.
\newblock Chatgpt in drug discovery: A case study on anticocaine addiction drug development with chatbots.
\newblock {\em Journal of Chemical Information and Modeling}, 63(22):7189--7209, 2023.

\bibitem{wang2020persistent}
R.~Wang, D.~D. Nguyen, and G.-W. Wei.
\newblock Persistent spectral graph.
\newblock {\em International journal for numerical methods in biomedical engineering}, 36(9):e3376, 2020.

\bibitem{wang2025large}
R.~Wang and T.~Schlick.
\newblock How large is the universe of rna-like motifs? a clustering analysis of rna graph motifs using topological descriptors.
\newblock {\em PLOS Computational Biology}, 21:e1013230, 2025.

\bibitem{wang2023persistent}
R.~Wang and G.-W. Wei.
\newblock Persistent path {L}aplacian.
\newblock {\em Foundations of Data Science}, 5(1):26--55, 2023.

\bibitem{wang2025join}
Y.~Wang, X.~Liu, Y.~Zhang, X.~Wang, and K.~Xia.
\newblock Join persistent homology (jph)-based machine learning for metalloprotein--ligand binding affinity prediction.
\newblock {\em Journal of Chemical Information and Modeling}, 65(6):2785--2793, 2025.

\bibitem{wee2023persistent}
J.~Wee, G.~Bianconi, and K.~Xia.
\newblock Persistent dirac for molecular representation.
\newblock {\em Scientific Reports}, 13(1):11183, 2023.

\bibitem{wee2024integration}
J.~Wee, J.~Chen, K.~Xia, and G.-W. Wei.
\newblock Integration of persistent {L}aplacian and pre-trained transformer for protein solubility changes upon mutation.
\newblock {\em Computers in Biology and Medicine}, page 107918, 2024.

\bibitem{wee2024evaluation}
J.~Wee and G.-W. Wei.
\newblock Evaluation of alphafold 3’s protein–protein complexes for predicting binding free energy changes upon mutation.
\newblock {\em Journal of Chemical Information and Modeling}, 64(16):6676--6683, 2024.

\bibitem{wee2025rapid}
J.~Wee and G.-W. Wei.
\newblock Rapid response to fast viral evolution using alphafold 3-assisted topological deep learning.
\newblock {\em Virus Evolution}, 11(1):veaf026, 2025.

\bibitem{wee2022persistent}
J.~Wee and K.~Xia.
\newblock Persistent spectral based ensemble learning ({PerSpect-EL}) for protein--protein binding affinity prediction.
\newblock {\em Briefings in Bioinformatics}, 23(2):bbac024, 2022.

\bibitem{wei2019protein}
G.-W. Wei.
\newblock Protein structure prediction beyond alphafold.
\newblock {\em Nature Machine Intelligence}, 1(8):336--337, 2019.

\bibitem{wei2022topological}
G.-W. Wei.
\newblock Topological {AI} forecasting of future dominating viral variants.
\newblock {\em SIAM News}, August, 09, 2022.

\bibitem{wei2023persistent2}
X.~Wei, J.~Chen, and G.-W. Wei.
\newblock Persistent topological laplacian analysis of {SARS-CoV}-2 variants.
\newblock {\em Journal of computational biophysics and chemistry}, 22(05):569--587, 2023.

\bibitem{wei2025persistent3}
X.~Wei and G.-W. Wei.
\newblock Persistent sheaf laplacians.
\newblock {\em Foundations of Data Science}, 7(2):446--463, 2025.

\bibitem{wei2025persistent}
X.~Wei and G.-W. Wei.
\newblock Persistent topological laplacians—a survey.
\newblock {\em Mathematics}, 13(2):208, 2025.

\bibitem{wu2018quantitative}
K.~Wu and G.-W. Wei.
\newblock Quantitative toxicity prediction using topology based multitask deep neural networks.
\newblock {\em Journal of chemical information and modeling}, 58(2):520--531, 2018.

\bibitem{wu2018topp}
K.~Wu, Z.~Zhao, R.~Wang, and G.-W. Wei.
\newblock Top{P--S}: {P}ersistent homology-based multi-task deep neural networks for simultaneous predictions of partition coefficient and aqueous solubility.
\newblock {\em Journal of computational chemistry}, 39(20):1444--1454, 2018.

\bibitem{xia2015fullerene}
K.~Xia, X.~Feng, Y.~Tong, and G.-W. Wei.
\newblock Persistent homology for the quantitative prediction of fullerene stability.
\newblock {\em Journal of computational chemistry}, 36(6):408--422, 2015.

\bibitem{xia2023persistent}
K.~Xia, X.~Liu, and J.~Wee.
\newblock Persistent {H}omology for {RNA Data A}nalysis.
\newblock In {\em Homology Modeling: Methods and Protocols}, pages 211--229. Springer, 2023.

\bibitem{xia2014persistent}
K.~Xia and G.-W. Wei.
\newblock Persistent homology analysis of protein structure, flexibility, and folding.
\newblock {\em International journal for numerical methods in biomedical engineering}, 30(8):814--844, 2014.

\bibitem{xia2015multidimensional}
K.~Xia and G.-W. Wei.
\newblock Multidimensional persistence in biomolecular data.
\newblock {\em Journal of computational chemistry}, 36(20):1502--1520, 2015.

\bibitem{xia2015persistent}
K.~Xia and G.-W. Wei.
\newblock Persistent topology for cryo-{EM} data analysis.
\newblock {\em International Journal for Numerical Methods in Biomedical Engineering}, 31(8), 2015.

\bibitem{xia2015multiresolution_large}
K.~Xia, Z.~Zhao, and G.-W. Wei.
\newblock Multiresolution persistent homology for excessively large biomolecular datasets.
\newblock {\em The Journal of chemical physics}, 143(13), 2015.

\bibitem{xia2015multiresolution}
K.~Xia, Z.~Zhao, and G.-W. Wei.
\newblock Multiresolution topological simplification.
\newblock {\em Journal of Computational Biology}, 22(9):887--891, 2015.

\bibitem{xu2024pld}
X.~Xu, C.~Wang, G.-W. Wei, and J.~Chen.
\newblock Pld-tree: Persistent laplacian decision tree for protein-protein binding free energy prediction.
\newblock {\em arXiv preprint arXiv:2412.18541}, 2024.

\bibitem{yang2024topological}
Y.~Yang, S.~Guo, S.~Li, Y.~Wu, and Z.~Qiao.
\newblock Topological data analysis combined with high-throughput computational screening of hydrophobic metal--organic frameworks: Application to the adsorptive separation of c3 components.
\newblock {\em Nanomaterials}, 14(3):298, 2024.

\bibitem{yao2009topological}
Y.~Yao, J.~Sun, X.~Huang, G.~R. Bowman, G.~Singh, M.~Lesnick, L.~J. Guibas, V.~S. Pande, and G.~Carlsson.
\newblock Topological methods for exploring low-density states in biomolecular folding pathways.
\newblock {\em The Journal of chemical physics}, 130(14), 2009.

\bibitem{zhang2024multi}
Y.~Zhang, C.~Shen, and K.~Xia.
\newblock Multi-cover persistence (mcp)-based machine learning for polymer property prediction.
\newblock {\em Briefings in Bioinformatics}, 25(6):bbae465, 2024.

\bibitem{zheng2023application}
S.~Zheng, H.~Ding, S.~Li, D.~Chen, and F.~Pan.
\newblock Application of topology-based structure feature for machine learning in material science.
\newblock {\em Chinese Journal of Structural Chemistry}, 42(7):100120, 2023.

\bibitem{zheng2025active}
S.~Zheng, X.-M. Zhang, H.-S. Liu, G.-H. Liang, S.-W. Zhang, W.~Zhang, B.~Wang, J.~Yang, X.~Jin, F.~Pan, et~al.
\newblock Active phase discovery in heterogeneous catalysis via topology-guided sampling and machine learning.
\newblock {\em Nature Communications}, 16(1):1--13, 2025.

\bibitem{zhu2023tidal}
Z.~Zhu, B.~Dou, Y.~Cao, J.~Jiang, Y.~Zhu, D.~Chen, H.~Feng, J.~Liu, B.~Zhang, T.~Zhou, and G.-W. Wei.
\newblock T{IDAL: Topology-Inferred Drug Addiction Learning}.
\newblock {\em Journal of Chemical Information and Modeling}, 63(5):1472--1489, 2023.

\bibitem{zia2024topological}
A.~Zia, A.~Khamis, J.~Nichols, U.~B. Tayab, Z.~Hayder, V.~Rolland, E.~Stone, and L.~Petersson.
\newblock Topological deep learning: a review of an emerging paradigm.
\newblock {\em Artificial Intelligence Review}, 57(4):77, 2024.

\bibitem{zia2025persistent}
M.~Zia, B.~Jones, H.~Feng, and G.-W. Wei.
\newblock Persistent directed flag laplacian (pdfl)-based machine learning for protein--ligand binding affinity prediction.
\newblock {\em Journal of Chemical Theory and Computation}, 21(8):4276--4285, 2025.

\bibitem{zomorodian2004computing}
A.~Zomorodian and G.~Carlsson.
\newblock Computing persistent homology.
\newblock In {\em Proceedings of the twentieth annual symposium on Computational geometry}, pages 347--356, 2004.

\end{thebibliography}
\bibliographystyle{abbrv}
\end{document}